\documentclass[twocolumn,preprintnumbers,amsmath,amssymb,pre]{revtex4}

\usepackage{graphicx}
\usepackage{color}
\usepackage{amsmath}

\begin{document}

\title{Quantum nuclear motion in silicene: Assessing structural and vibrational
properties through path-integral simulations}
\author{Carlos P. Herrero}
\author{Miguel del~Canizo}
\affiliation{Instituto de Ciencia de Materiales de Madrid,
         Consejo Superior de Investigaciones Cient\'ificas (CSIC),
         Campus de Cantoblanco, 28049 Madrid, Spain }
\date{\today}

\begin{abstract}
This paper explores the interplay between quantum nuclear motion
and anharmonicity, which causes nontrivial effects on the structural
and dynamical characteristics of silicene, a two-dimensional (2D)
allotrope of silicon with interesting electronic and mechanical properties. 
Employing path-integral molecular dynamics (PIMD) simulations, we investigate 
the quantum delocalization of nuclei, unraveling its impact on the behavior 
of silicene at the atomic scale. Our study reveals that this delocalization
induces significant deviations in the structural parameters of silicene, 
influencing in-plane surface area, bond lengths, angles, compressibility,
and overall lattice dynamics. 
Through extensive simulations, we delve into the temperature-dependent 
behavior between 25 and 1200~K, unveiling the role of quantum nuclear 
fluctuations in dictating thermal expansion and phonon spectra.
The extent of nuclear quantum effects is assessed by comparing results
of PIMD simulations using an efficient tight-binding Hamiltonian,
with those obtained from classical molecular dynamics simulations.
The observed quantum effects showcase non-negligible deviations 
from classical predictions, emphasizing the need for accurate quantum 
treatments in understanding the material's behavior at finite temperatures. 
At low $T$, the 2D compression modulus of silicene decreases by a 14\% due
to quantum nuclear motion.
We compare the magnitude of quantum effects in this material
with those in other related 2D crystalline solids, such as graphene and 
SiC monolayers.   \\

\end{abstract}

\maketitle

\section{Introduction}

Silicene, a two-dimensional allotrope of silicon, has garnered significant 
attention due to its relevant properties and potential applications 
across diverse fields. Serving as the silicon counterpart to graphene, 
silicene boasts advantages rooted in its atomic structure and electronic 
properties \cite{si-za14,si-ta15,si-gr16,si-mo18,si-zh15,si-kh20,
si-ta21,si-gh23}.  Its intrinsic semiconducting nature and compatibility 
with existing silicon-based technology position it as a promising 
candidate for next-generation electronics. With remarkably high 
electronic mobility and a tunable bandgap, silicene paves the way 
for efficient and high-speed transistors, enhancing the capabilities 
of integrated circuits.
Beyond electronics, silicene's versatility shines through. Its flexibility, 
interaction with light, and compatibility with various substrates facilitate 
applications in sensing, photonics, and nanoelectromechanical systems 
\cite{si-ni12,si-gu15b,si-zh16,si-mo18,si-gu21,si-gu21b,si-do19}.
This breadth of potential applications underscores the importance of further 
exploration into silicene's properties and functionalities.

From a theoretical standpoint, silicene has been investigated 
using \textit{ab-initio} density-functional theory (DFT) calculations, 
which have provided reliable insights into its properties 
\cite{si-ta21,si-ca09,si-tr14}. 
Unlike graphene's planar structure, these calculations reveal that silicene 
tends to adopt a buckled configuration, indicating a competition between 
$sp^2$ and $sp^3$ electronic hybridization, with the former prevailing 
in graphene \cite{si-ta21,si-ca09,si-tr14}.
DFT-based analyses have been instrumental in examining various aspects 
of silicene's behavior. They have shed light on its bending characteristics 
\cite{si-ku20,si-ba16,si-pi19}, mechanical properties 
\cite{si-zh12,si-mo17,si-yo21,si-ze18}, phonon behavior 
\cite{si-hu15,si-ya13,si-pe16b}, and thermal conductivity 
\cite{si-pe16,si-xi14,si-gu15}. Moreover, DFT calculations have 
extended to the study of silicene multilayers, elucidating the evolution 
of physical properties from single-layer configurations to bulk silicon 
\cite{si-ja21,si-pa15,si-ip22,si-ya14b}. 

Finite-temperature atomistic simulations, notably molecular dynamics, 
have been utilized by diverse research teams to examine
the mechanical and thermal characteristics of both free-standing 
and supported silicene, operating within both equilibrium and 
nonequilibrium scenarios. These simulations often relied on effective 
potentials such as Tersoff, Stillinger-Weber, and modified embedded 
atom methods \cite{si-be14,si-da18,si-hu13b,si-in11,si-lo23, 
si-mi18,si-pe14,si-ro19, si-wa15,si-ro17}.
The breadth of these investigations encompasses a wide range of silicene 
properties, including tensile strength, bending behavior, oscillatory 
dynamics, and equilibrium responses under various loading conditions. 
Moreover, researchers have scrutinized the effects of vacancies, 
the evolution of fracture, and thermal conductivity, frequently drawing 
comparisons with graphene for a comprehensive understanding of silicene's 
characteristics \cite{si-be14,si-da18,si-hu13b,si-in11,si-lo23, 
si-mi18,si-pe14,si-ro19, si-wa15,si-ro17}.

Despite the wealth of dependable insights offered by classical molecular 
dynamics (MD) simulations, the intrinsic quantum nature of silicon nuclei 
introduces potential discrepancies in calculated physical properties, 
particularly at temperatures below the Debye temperature of silicene, 
$\Theta_D$. The range of estimates for $\Theta_D$ falls 
between 550 and 680 K \cite{si-wa15,si-ya14b,si-pe16b}, notably 
surpassing typical room temperatures.
To reconcile this quantum aspect, the Feynman path-integral method emerges 
as a suitable approach, leveraging MD or Monte Carlo 
sampling techniques \cite{fe72,gi88,he14,ce95}. By employing the
path-integral procedure,
the nuclear degrees of freedom can be effectively quantized, thus 
accommodating thermal and quantum fluctuations at finite temperatures. 
This framework facilitates quantitative analyses of anharmonic effects 
in condensed matter, particularly in the realm of two-dimensional (2D) 
materials.
In this regard, investigations into quantum nuclear motion have unveiled 
significant impacts on the structural and thermodynamic properties of 
various 2D materials, including graphene \cite{he16,br15}, as well as 
monolayers of SiC and BN \cite{he22,br22}. 

In the low-temperature limit, graphene and 2D SiC exhibit quantum nuclear 
motion, resulting in zero-point energies of 172 and 86 meV/atom, 
respectively. For these layered materials, nuclear quantum effects 
significantly impact the interatomic distances, layer areas, and 
thermodynamic properties at low temperatures, and can even be noticeable 
at room temperature \cite{br15,he16,he22}.
In the paradigmatic case of graphene, zero-point expansion due to quantum
nuclear motion leads to an approximate 1\% increase in the in-plane area. 
Additionally, the temperature dependence of the in-plane area can vary 
qualitatively when derived from classical versus path-integral 
molecular dynamics (PIMD) simulations, even at temperatures between 
$T = 300$ and 1000~K \cite{br15,he16}.
At low temperatures, PIMD simulations reveal that the specific heat for 
stress-free graphene changes as $c_p \sim T$, transitioning to a 
$c_p \sim T^2$ dependence under tensile stress \cite{he18b}. 
Such quantum simulations have also been utilized to study isotopic 
effects in this material \cite{he20c}.
For 2D layers of SiC, the bending constant and 2D modulus of compression 
are lower than those of graphene. This indicates that SiC is more susceptible 
to in-plane stresses. Consequently, nuclear quantum motion, combined 
with anharmonicity of vibrational modes, produces more directly 
observable effects in SiC layers.

For silicene, a deep understanding of its structural, mechanical, and 
thermodynamic properties is essential for an accurate characterization, 
especially considering its potential technological applications. 
These properties are expected to be influenced by the quantum motion of 
silicon nuclei, similar to graphene and SiC layers. However, a key 
difference between silicene and these other 2D materials is its lack 
of strict planarity, which may significantly affect the elastic and 
electronic properties. Consequently, the coupling of electronic and 
nuclear degrees of freedom (i.e., electron-phonon interaction) in the 
presence of competing $sp^2$ and $sp^3$ hybridizations is likely 
crucial for an accurate characterization of silicene at finite temperatures.

In this paper, we present the outcomes of comprehensive 
PIMD simulations conducted on silicene layers. 
By leveraging this theoretical framework, we embark 
on a quantitative exploration of thermal and quantum 
nuclear fluctuations spanning a wide range of temperatures. 
Our objective is to furnish an in-depth analysis, informed by analogous 
research on other 2D materials, thereby offering both qualitative and 
quantitative insights that significantly enrich our comprehension of 
these systems.

In our study, we delve into the influence of quantum nuclear motion 
on both the structural and vibrational characteristics of silicene, 
contrasting these effects with the expected behavior under classical 
motion (i.e., classical dynamics of the Si atoms). 
Our simulations employ interatomic interactions derived from an effective 
tight-binding (TB) Hamiltonian calibrated to DFT-derived data. 
We scrutinize the impact of anharmonicity on the physical attributes 
of silicene by juxtaposing our simulation outcomes with those obtained 
from a harmonic approximation for the vibrational modes.
The realm of path-integral simulations concerning Si-containing materials 
has seen prior exploration, primarily focusing on aspects such as atomic 
quantum delocalization and anharmonic effects \cite{no96,he22}. 

The paper is structured as follows. In Sec.~II, we detail 
the computational methodologies underpinning our calculations, 
encompassing the tight-binding method and PIMD.
Sec.~III elaborates on the harmonic approximation used for studying 
the vibrational density of states, important for analyzing 
anharmonicities in silicene. Moving to Sec.~IV, we delve into an analysis 
of the internal energy, dissecting its kinetic and potential components.
Sec.~V is dedicated to examining structural properties, including interatomic 
distance, Si-Si-Si angle, and in-plane area.
Sec.~VI offers insights into the 2D modulus of compression 
at various temperatures. Finally, 
the main results are summarized in Sec.~VII.

\section{Simulation method}

Molecular dynamics and Monte Carlo simulations, employing the Feynman 
path-integral formulation of quantum statistical mechanics, stand as 
pivotal tools for analyzing many-body systems at finite 
temperatures. Among these techniques, PIMD simulations 
are renowned for their adeptness in capturing quantum phenomena 
inherent in atomic nuclei behavior. This methodology hinges on an 
isomorphism which equates a quantum system to a virtual classical one.
In this approach, each quantum particle, such as a Si nucleus in our context, 
is symbolized by a ring polymer comprising $N_{\rm Tr}$ (Trotter Number) 
beads interconnected by harmonic strings \cite{fe72,gi88,ce95}. 
Through this association, the partition function of the quantum 
system assumes a structure reminiscent of a classical system. In this 
framework, an effective potential emerges from the harmonic interactions 
among neighboring beads within each of the $N$ (number of particles) 
polymer rings. The formal exactitude of the partition function is attained 
for $N_{\rm Tr} \to \infty$. 
Additional insights into this computational method can be 
found elsewhere \cite{gi88,ce95,he14,ca17}.

In our simulations of silicene, the dynamics of atomic nuclei 
are governed by a Born-Oppenheimer energy surface derived from an effective 
non-orthogonal tight-binding Hamiltonian \cite{po95,gu96,go97}, 
constructed via DFT calculations. 
This Hamiltonian emerges as a balance between 
accuracy, transferability, and computational efficiency.
The approach presented by Porezag et al.~\cite{po95} operates on 
the basis of the linear combination of atomic orbitals (LCAO) 
method within the local density approximation (LDA). It involves 
the creation of a pseudo-atom endowed with a collective confinement 
potential derived from the covalent radius of the target atom. 
The eigenfunctions of this pseudo-atom serve as the basis for a minimal 
orbital set, where the wavefunctions of the extended system are expressed 
through an LCAO equation. The computation of Hamiltonian and overlap 
matrices within the LDA employs a two-center approximation and Slater-Koster 
integrals. This hybrid {\em ab initio}-parameterized 
TB methodology harnesses appropriate input densities and potentials, 
without the need for experimental data adjustments.
Such a TB model, characterized by its integration of {\em ab initio} 
calculations with parameterized elements, has found prior application 
in studies concerning silicon \cite{fr95,si-kl99,si-kl99b,sc-ka05} and 
Si-containing materials \cite{sc-sh01,gu96,ra08,he22}.
Other TB parameterizations have been also employed to study several properties
of silicon by MD simulations, including defect diffusion and
phase transformations \cite{si-ma02,si-al01,si-co05}.

We performed PIMD simulations on silicene layers within the isothermal-isobaric 
ensemble, keeping the number of particles ($N$), temperature ($T$), and 
in-plane stress ($\tau$) constant. The stress, measured as force per unit length, 
corresponds to what is referred to in the literature as mechanical 
or frame tension \cite{ra17,fo08,sh16}. The bead coordinates within the ring 
polymers were represented using staging variables. We integrated the equations 
of motion, specifically adapted for 2D materials, using reversible 
integrators with factorization techniques for the Liouville time-evolution 
operator. In particular, we utilized the reversible system propagator 
algorithm (RESPA) \cite{ma96,ra20}. This approach allowed for the adaptive 
use of distinct time steps: $\Delta t = 1$~fs for dynamical variables 
associated with interatomic forces, and $\delta t = 0.25$~fs for faster 
dynamical variables involving harmonic bead interactions and thermostats.

Temperature control was achieved by coupling a chain of four Nos\'e-Hoover 
thermostats, with a ``mass'' parameter $Q = \beta \hbar^2 / 5 N_{\rm Tr}$, 
to each staging variable ($\beta = 1 / k_B T$ and $k_B$ is Boltzmann's 
constant). Consequently, each of the $3N$ atomic momentum coordinates was 
coupled to a separate chain of $M$ thermostats. This extensive thermostatting 
of the system is essential in PIMD simulations to prevent the emergence of 
ergodicity issues. In classical MD simulations, it would also be appropriate 
to use a single thermostat chain for all atoms. The primary reason for 
employing massive thermostatting in our calculations is that the additional 
computational cost is minimal, allowing the same computational code to 
be used for both PIMD and classical MD simulations.

To maintain a constant in-plane stress ($\tau = 0$ in this case),
we coupled the $xy$ surface area of the simulation cell to a barostat
with an effective ``mass'' $W$ (with dimensions of energy $\times$ time$^2$),
which is further connected to a chain of $M$ thermostats \cite{tu10,he14}.
Specifically, for the time derivative of the in-plane area of the simulation
cell, $A = N A_p$, the equation $dA / dt = 2 A \, p_A / W$ was applied,
where $p_A$ is an effective ``momentum'' associated with $A$, having
dimensions of energy $\times$ time.
In this work, we set $M = 4$ and $W = 3.2 \times 10^{15}$ eV~fs$^2$,
following Ref.~\cite{ra20}. We verified that this choice of parameters is
sufficient for ensuring controlled system equilibration, as well as for
maintaining the desired temperature and external stress. In particular,
using $M = 4$ ensures rapid equilibration, and no significant change was
observed with larger values of $M$.

The dynamical equations in the $N \tau T$ ensemble involve the time 
derivatives of the position and momentum coordinates of the dynamic 
degrees of freedom in the extended system, which includes staging modes, 
chains of Nos\'e-Hoover thermostats, the volume, and the barostat 
\cite{tu10,he14}. The thermostats and barostat introduce friction terms 
into the dynamical equations, resulting in non-Hamiltonian dynamics. 
However, it is possible to define a quantity analogous to the total energy 
in the extended system, which is conserved during the dynamical process. 
This conserved quantity serves as a useful check for the accuracy of the 
numerical integration of the equations of motion 
(see, e.g., Ref.~\cite{tu10}).
To calculate the kinetic energy $E_{\rm kin}$, we used the virial 
estimator. This method is particularly advantageous due to its lower 
statistical uncertainty compared to that of the potential energy, 
especially at high temperatures \cite{tu10,he82}.

The choice of $N_{\rm Tr}$ in PIMD simulations is influenced by two main 
considerations. Firstly, it is important to ensure consistent accuracy 
across the entire temperature range. To achieve this, we scale the Trotter 
number according to the relationship $N_{\rm Tr} T = 6000$~K, following 
earlier guidelines outlined in Refs.~\cite{he16,sc-he24}. 
At room temperature ($T = 300$~K), this corresponds to ring polymers 
with 20 beads.
Secondly, a finite $N_{\rm Tr}$ introduces a cutoff frequency, which in our 
case is $\omega_c = N_{\rm Tr} k_B T / \hbar \approx$ 4200~cm$^{-1}$.
This frequency significantly exceeds the vibrational frequencies observed 
in silicene. Thus, the highest frequency of optical phonons 
is $\omega_m \approx$ 680~cm$^{-1}$, resulting in 
$\omega_c / \omega_m \approx 6$.
This ratio ensures good convergence of vibrational properties with respect 
to the Trotter number in our quantum simulations.
We verified this convergence at $T = 300$~K by considering values of 
$N_{\rm Tr}$ up to 60, and found that the results for the variables 
studied here were consistent within statistical error bars when compared 
to $N_{\rm Tr}$ = 20. Specifically, the energy differences were smaller 
than 1~meV/atom.
With our choice of $N_{\rm Tr}$ for each temperature $T$, the imaginary time 
step between successive beads in the ring polymers is given by 
$\beta \hbar / N_{\rm Tr}$ = 1.3~fs, which is comparable to the time step 
used for integrating the dynamical equations in our PIMD simulations 
($\Delta t$ = 1~fs, see above).

We used rectangular simulation cells with similar side lengths in the 
$x$ and $y$ directions of the $xy$ reference plane, applying periodic 
boundary conditions. Silicon atoms were allowed unrestricted movement 
in the out-of-plane direction, implementing free boundary conditions along 
the $z$ axis to model a free-standing silicene layer.
We verified that isotropic changes in cell dimensions during $N \tau T$ 
simulations yielded the same results for the studied variables as using 
flexible cells, which allow independent changes in the $x$ and $y$ axis 
lengths, along with deformations of the rectangular shape. 
Additionally, to assess the impact of boundary conditions, we utilized 
supercells of the primitive hexagonal cell and observed no changes 
in our results.

In our simulations, we considered silicene supercells containing $N =$ 60, 
112, and 216 atoms. To efficiently sample the electronic degrees of freedom 
in reciprocal space, we focused solely on the $\Gamma$ point (${\bf k} = 0$). 
While considering larger ${\bf k}$ sets may induce slight changes in 
total energy, these variations are negligible for the energy differences 
pertinent to our discussion. We have observed that any shift in the minimum 
energy $E_0$ becomes increasingly inconsequential with larger cell sizes.
Sampling the configuration space encompassed temperatures ranging from 
25 to 1200 K. Typical simulation protocols involved $2 \times 10^5$ PIMD 
steps for system equilibration, followed by $8 \times 10^6$ steps for 
computing average properties. 
We have checked that this simulation length is much larger than the 
autocorrelation time $\Omega$ of the variables studied in this paper, 
for our zero-stress conditions. In particular, $\Omega$ for the in-plane 
area $A_p$ appreciably increases as the system size rises, and 
for the largest size considered here, we have found autocorrelation 
times shorter than $10^5$ simulation steps.

\begin{figure}
\vspace{-7mm}
\includegraphics[width=7cm]{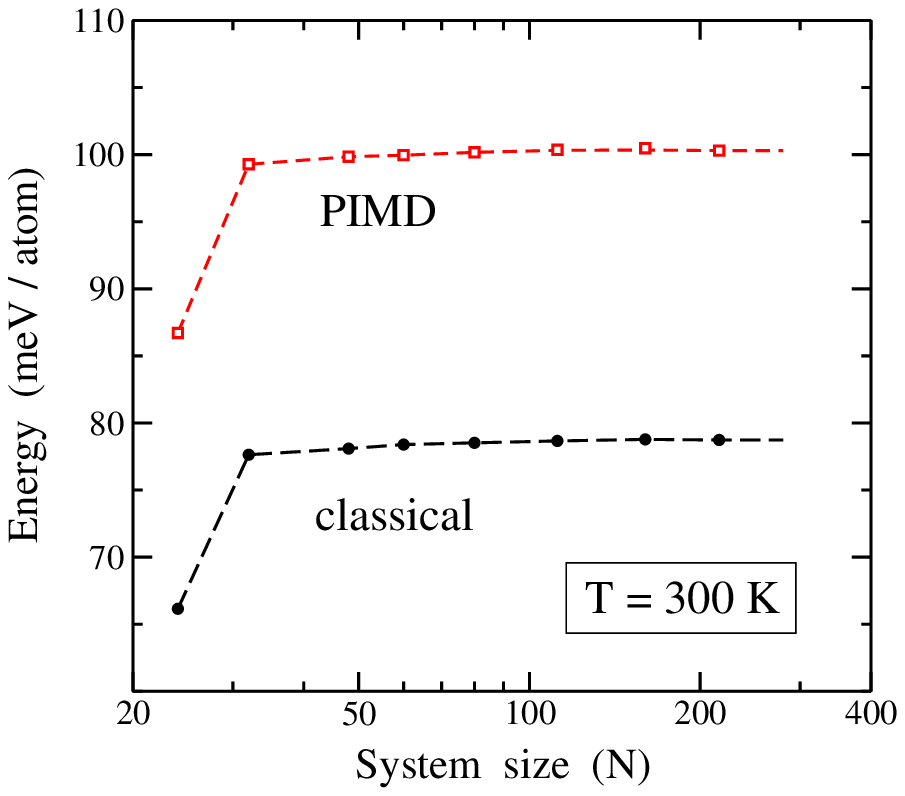}
\vspace{-5mm}
\caption{System size dependence of the energy for silicene at $T$ = 300~K,
computed using the TB Hamiltonian presented in this study. Circles represent
results from classical MD, while open squares depict outcomes from PIMD
simulations.  Lines are guides to the eye.
Energy values are referenced to the minimum-energy state for a system
size $N = 216$.
}
\label{f1}
\end{figure}

To gauge the significance of quantum effects elucidated by our PIMD 
simulations, we conducted classical MD simulations using
the same TB Hamiltonian. In this scenario, $N_{\rm Tr} = 1$, 
resulting in the collapse of ring polymers into single beads. 
This approach allows for a direct comparison between quantum and 
classical data, shedding light on the extent to which quantum 
fluctuations influence the behavior of the system.

In Fig.~1, we depict the size dependency of the energy derived from 
our simulations of silicene conducted at $T$ = 300~K. Circles represent 
data from classical MD simulations, while open squares denote results 
obtained from PIMD simulations. Across both datasets, we observe 
a slow increase in energy for $N > 30$. Energy changes
for varying system sizes remain below 2~meV for $N > 50$, 
and decrease to less than 1~meV for $N > 100$.
It is worth noting that the disparity between the two datasets remains 
constant at 22~meV/atom for $N > 30$. 

Furthermore, to contextualize our findings concerning buckled silicene, 
the predicted configuration for this 2D material, we conducted classical 
MD simulations of planar silicene, characterized by a strictly flat 
configuration. This planar structure has been previously scrutinized 
by several researchers employing diverse theoretical methodologies 
\cite{si-ca09,si-sa09,si-sc13,si-gr16,si-mo18,si-ja21,si-wa15}. 
Comparing results for both buckled and planar silicene offers 
insights into the energetics and dynamics of this material, 
enriching our understanding of its fundamental properties.

\begin{figure}
\vspace{-7mm}
\includegraphics[width=7cm]{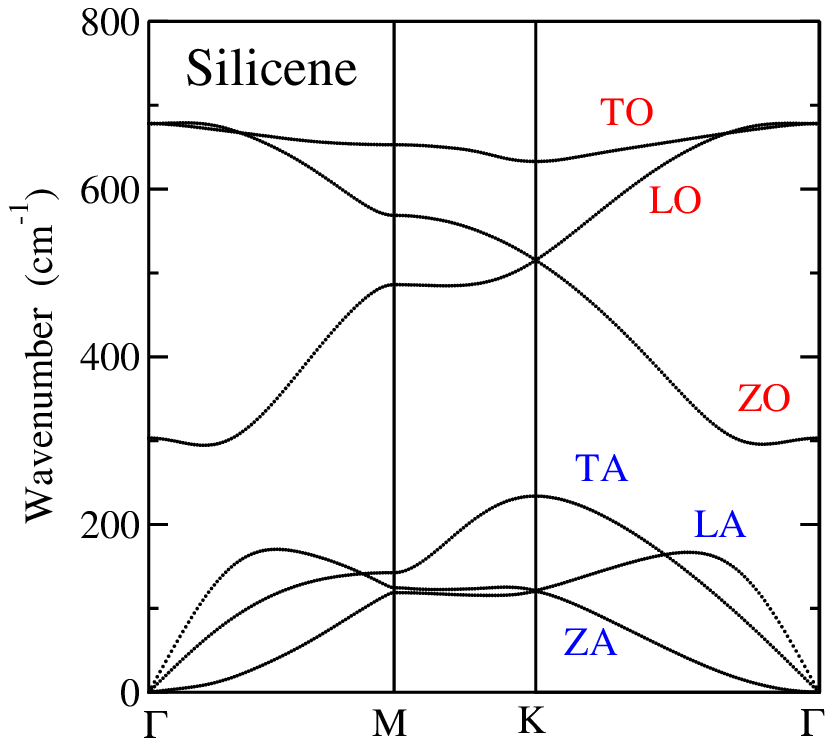}
\vspace{-5mm}
\caption{Phonon dispersion branches of silicene, $\omega_j({\bf k})$,
found from diagonalization of the dynamical matrix for the
TB Hamiltonian used in this work.
Labels indicate the six phonon bands.
}
\label{f2}
\end{figure}

\section{Harmonic approximation}

In evaluating the significance of anharmonicity in our PIMD simulations 
for silicene, we adopt a harmonic approximation (HA) for the vibrational 
modes. In this method, frequencies are derived for the minimum-energy 
configuration and are assumed to remain constant, irrespective of 
temperature. Similarly, the in-plane area, $A_p$, is treated as constant, 
thereby disregarding thermal fluctuations, as well as thermal 
expansion or contraction.
While the HA typically provides reliable estimates at low temperatures 
in solids, its accuracy diminishes as temperature increases due to 
the growing influence of anharmonicity. Consequently, one anticipates 
an increasing disparity between the predictions of the HA and the more 
accurate simulation results as temperature rises.

\begin{figure}
\vspace{-15mm}
\includegraphics[width=7cm]{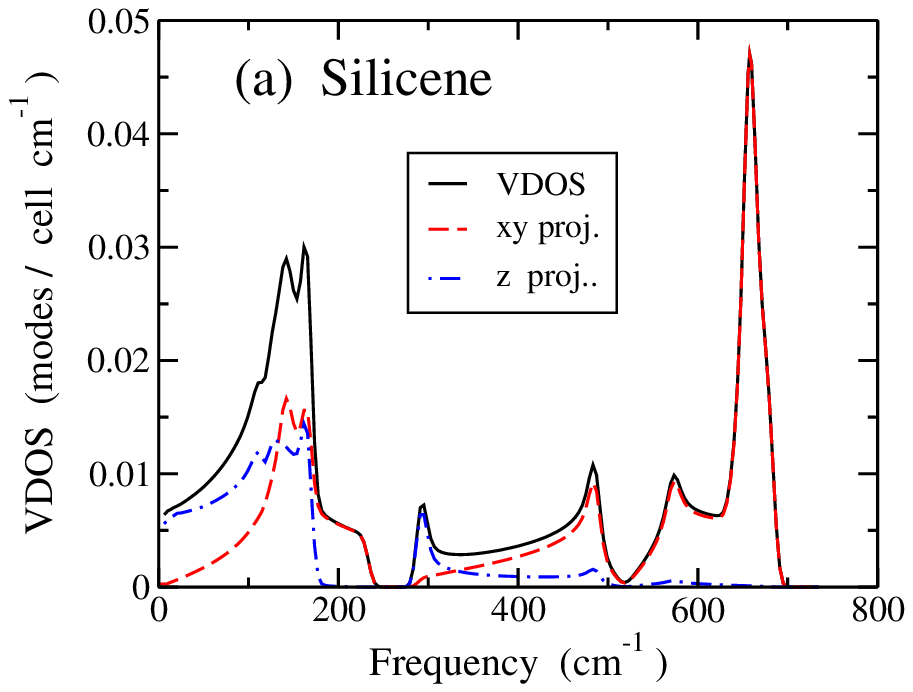}
\includegraphics[width=7cm]{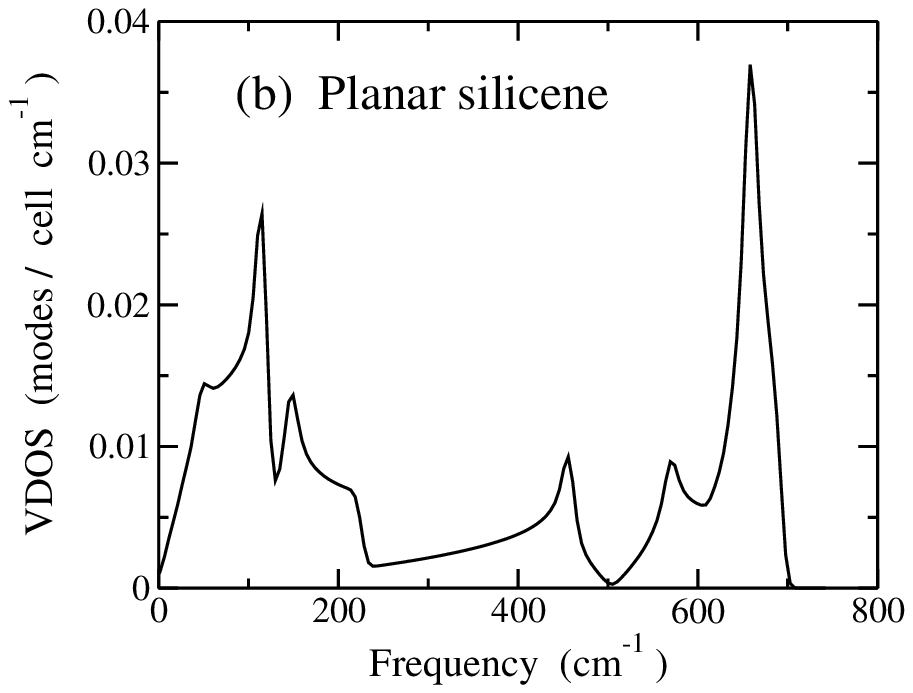}
\vspace{-5mm}
\caption{(a) Vibrational density of states of silicene,
obtained in the HA for the TB Hamiltonian used in this paper (solid
line). Dashed and dashed-dotted lines indicate the in-plane
($xy$) and out-of-plane ($z$) projections of the VDOS, respectively.
(b) VDOS for planar silicene (only $xy$ component).
}
\label{f3}
\end{figure}

In Fig.~2, we illustrate the phonon bands of silicene along symmetry 
directions within the 2D hexagonal Brillouin zone, as determined in 
the HA through the diagonalization of 
the dynamical matrix, employing the TB Hamiltonian. 
Labels denote the six phonon branches depicted in the plot.
The phonon dispersion showcased in Fig.~2 is similar to those
observed in studies utilizing empirical potentials \cite{si-zh14} and 
DFT calculations 
\cite{si-hu15,si-sc13,si-ya13,si-gu15,si-pe16b,si-ze18}. 
Notably, we observe the presence of the flexural ZA band, characteristic 
of 2D materials.

To directly quantify the overall anharmonicity, we computed the 
vibrational density of states (VDOS), $\rho(\omega)$, across the entire 
Brillouin zone. Employing numerical integration, we followed 
the methodology outlined in Ref.~\cite{ra86}. Fig.~3(a) illustrates 
the resulting VDOS for silicene, depicted by a solid line.

For a more thorough analysis, we computed the projection of the eigenvectors 
resulting from the diagonalization of the dynamical matrix onto 
the in-plane $x$, $y$, and out-of-plane $z$ directions, 
thereby determining their respective 
contributions to the VDOS using the expression \cite{si-to23}:
\begin{equation}
  \rho^j(\omega, \hat{\mathbf{n}}) = \frac{1}{N}
     \sum_\lambda \delta(\omega - \omega_\lambda)
        | \hat{\mathbf{n}} \cdot \mathbf{e}^j_\lambda |^2  \; ,
\end{equation}
where $j$ denotes the atoms within the crystallographic unit cell 
(here, $j = 1, 2$), $\hat{\mathbf{n}}$ represents a unit projection vector, 
$\lambda$ iterates over the frequency eigenvalues, and 
$\mathbf{e}^j_\lambda$ signifies the eigenvector of mode $\lambda$. 
Subsequently, for a given direction, let us say $x$, the VDOS along 
that direction is given by 
$\rho_x(\omega) = \rho^1(\omega, \hat{\mathbf{n}}_x) + 
\rho^2(\omega, \hat{\mathbf{n}}_x)$.

In Fig.~3(a), we have plotted the projected VDOS along the $z$ direction 
(dashed-dotted line), as well as the projection onto the layer 
plane encompassing the $x$ and $y$ directions (dashed line). 
The predominant contribution of the $z$ component is 
evident for frequencies below 200~cm$^{-1}$, corresponding to acoustic 
modes. 

Our calculations on silicene employing the TB Hamiltonian reveal a distinct 
gap in the VDOS between 240 and 280~cm$^{-1}$, 
delineating the frequency regimes of acoustic and optical modes. 
This stands in contrast to graphene, where such a gap is absent 
\cite{wi04,mo05,ya08,ra19}, and its presence in silicene 
can be attributed to the lack of planarity.
To elucidate this disparity, we computed the VDOS for planar silicene, 
as depicted in Fig.~3(b). Since planar silicene does not represent 
the energetically favorable configuration under the employed TB Hamiltonian, 
computing vibrational modes in three-dimensional space results in 
unphysical imaginary frequencies for out-of-plane ($z$ direction) vibrations. 
Therefore, the VDOS shown in Fig.~3(b) captures the dynamics 
of strictly planar silicene in 2D space, encompassing only 
in-plane vibrations.
Above 300~cm$^{-1}$, this VDOS closely resembles that 
of buckled silicene, as depicted in Fig.~3(a), since in this frequency 
range it is predominantly influenced by projections in the $xy$ plane. 
Notably, the gap observed between 240 and 280~cm$^{-1}$ in the buckled 
material is absent in the planar counterpart.

\begin{figure}
\vspace{-7mm}
\includegraphics[width=7cm]{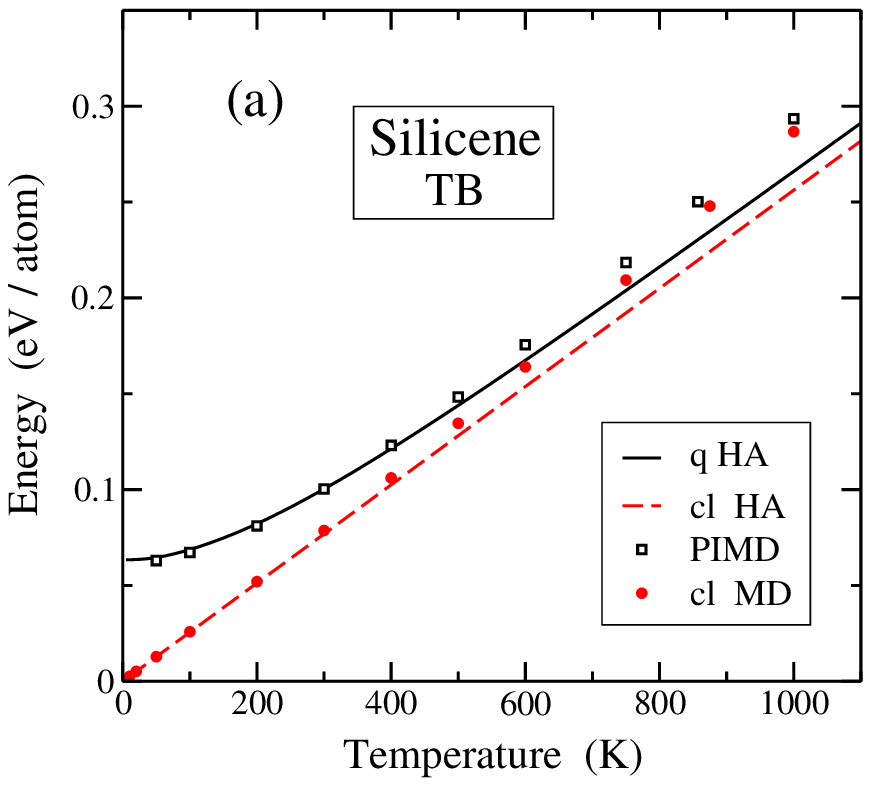}
\includegraphics[width=7cm]{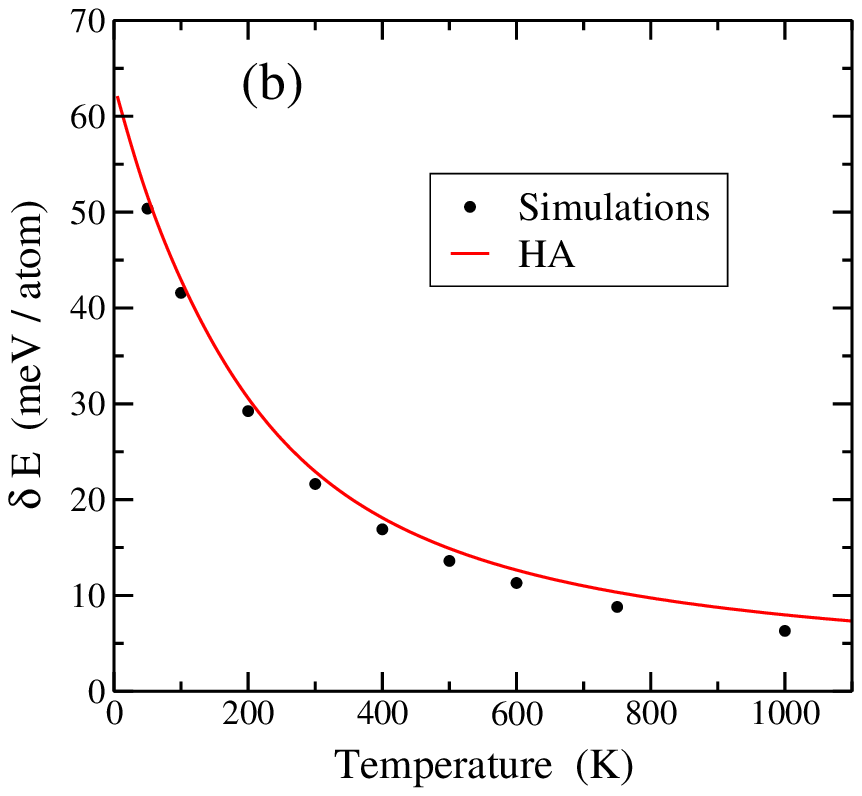}
\vspace{-5mm}
\caption{(a) Energy as a function of temperature. Symbols indicate
results of simulations using the TB Hamiltonian: classical
MD (circles) and PIMD (open squares).
The solid line is the vibrational energy derived from the
VDOS shown in Fig.~3(a).
The dashed line indicates the classical vibrational energy
in a harmonic approximation: $E_{\rm vib}^{\rm cl} = 3 k_B T$.
(b) Difference $\delta E$ between quantum and classical results.
Solid circles, data derived from simulations; continuous line, harmonic
approximation.
}
\label{f4}
\end{figure}

\section{Energy}

In this section, we examine the evolution of silicene's energy with 
respect to temperature, as derived from our PIMD simulations.
We juxtapose these findings with results from classical MD simulations 
and the HA, aiming to discern the effects of quantum nuclear motion
and anharmonicity of the atomic vibrations.

In Fig.~4(a), we present the temperature-dependent energy, $E - E_0$, 
extracted from PIMD simulations (open squares), where $E_0$ denotes the 
energy of the minimum-energy configuration. This configuration corresponds 
to the classical zero-temperature limit, devoid of quantum delocalization 
effects. For comparison, the results obtained from classical MD 
simulations are depicted as circles.
At low temperatures, the quantum data converge to a zero-point energy, 
$E_{\rm ZP} =$ 62 meV/atom. Converted to a mean frequency, 
$\overline{\omega}$, using the relationship 
$E_{\rm ZP} = \hbar \, \overline{\omega} / 2$, this yields 
$\overline{\omega} =$ 333 cm$^{-1}$. Remarkably, this value closely 
aligns with the average frequency derived from the VDOS showcased in 
Fig.~3(a), which amounts to $\overline{\omega} =$ 340 cm$^{-1}$.

For comparison with the simulation data, we present results of 
the HA in both classical (dashed line) and 
quantum (solid line) scenarios. At temperature $T$, the quantum-mechanical 
vibrational energy per atom, $E_{\rm vib}$, within the HA framework is 
determined from the VDOS $\rho(\omega)$ through a continuous 
approximation, given by:
\begin{equation}
 E_{\rm vib} = \frac14 \int_0^{\omega_m} \hbar \, \omega
     \coth \left( \frac12 \beta \hbar \, \omega \right)
      \rho(\omega) d \omega  \, ,
\label{evib2}
\end{equation}
where $\beta = 1 / k_B T$ 
and $\omega_m$ denotes the maximum frequency in the phonon spectrum. 
To account for the six degrees of freedom in a crystallographic unit cell 
(comprising two Si atoms), we ensure the normalization condition:
\begin{equation}
  \int_0^{\omega_m}  \rho(\omega)  d \omega  = 6   \, .
\label{intg}
\end{equation}

At low temperatures, the classical simulation data closely adhere 
to the harmonic approximation ($E = E_0 + 3 k_B T$), with both datasets 
exhibiting deviations from each other for $T > 400$~K. 
This deviation progressively increases to reach 12\% (30 meV/atom) 
at $T = 1000$~K.
We note that the departure of the classical energy 
from the harmonic expectation is solely attributed to the potential 
energy. In this scenario, the kinetic energy per atom remains constant, 
given by $E_{\rm kin} = 3 k_B T / 2$, regardless of anharmonicity.
The divergence between the PIMD results and the quantum HA
becomes noticeable at approximately the same temperature 
of 400~K. By $T = 1000$~K, this deviation reaches 10\% (27~meV/atom), 
after which classical and quantum simulation results demonstrate 
convergence with increasing temperature.

\begin{figure}
\vspace{-7mm}
\includegraphics[width=7cm]{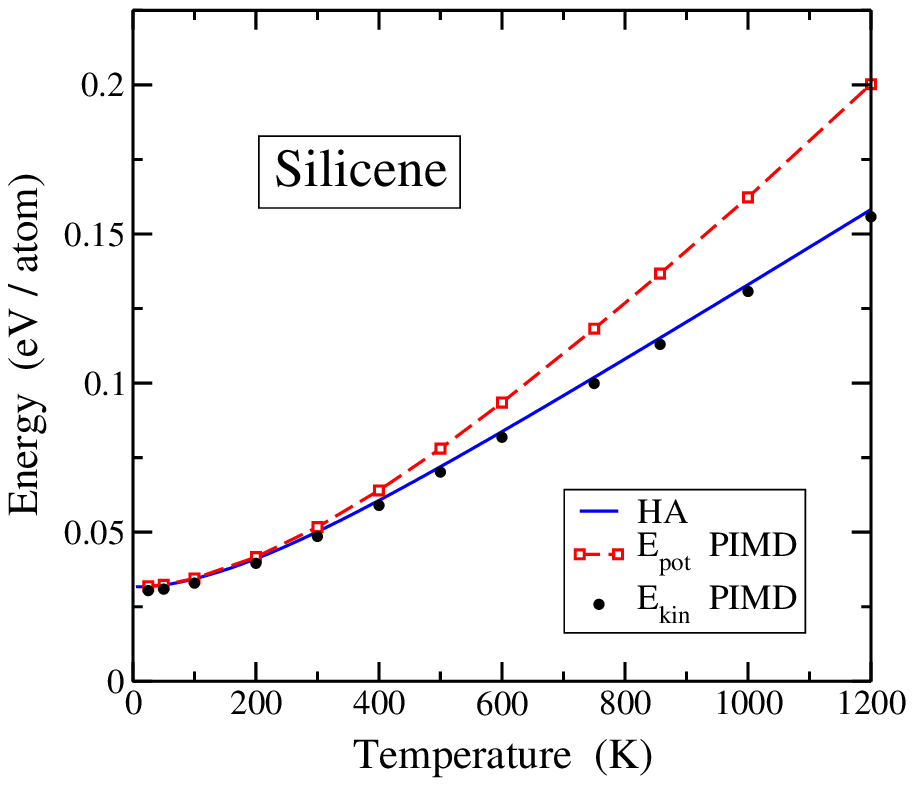}
\vspace{-5mm}
\caption{Temperature dependence of the kinetic (solid circles) and
potential energy (open squares), derived from PIMD simulations.
The solid line represents the kinetic and potential energy
derived from the HA. The dashed curve is a guide to the eye.
}
\label{f5}
\end{figure}

The discrepancy between quantum and classical outcomes for the energy, 
$\delta E$, spanning from $T = 0$ to 1000 K, is depicted 
in Fig.~4(b). This plot illustrates the data obtained from our simulations 
(circles), alongside the energy difference derived from the HA (solid line).
At low temperatures, the difference $\delta E$ derived from the simulations 
appears marginally smaller than the discrepancy inferred from the HA. 
However, as the temperature escalates, this disparity gradually expands.

A criterion for gauging the extent of overall anharmonicity lies 
in comparing the kinetic and potential energies, which are separately 
provided by PIMD simulations. 
In the realm of harmonic vibrations, both energies coincide at all 
temperatures, as dictated by the virial theorem, irrespective of 
the underlying approach, classical or quantum.
Fig.~5 illustrates the kinetic and potential energies derived from 
PIMD simulations as a function of temperature: $E_{\rm pot}$ 
(depicted by open squares) and $E_{\rm kin}$ (represented by circles). 
The solid curve portrays the outcome of the HA, 
derived from the VDOS using Eq.~(\ref{evib2}), 
where both energies align. The dashed line serves as a visual guide 
for the potential energy.
For temperatures below 400~K, the potential energy closely mirrors 
the kinetic energy, while the former escalates more rapidly than the 
latter at higher $T$. Anharmonicities in the interatomic potential 
lead to a notable increase in $E_{\rm pot}$ with temperature, 
while the kinetic energy remains closely aligned with the harmonic 
approximation. At $T = 1000$~K, the difference
$E_{\rm pot} - E_{\rm kin}$ amounts to 24\% of the kinetic energy.

Our findings contribute to a deeper understanding of electron-phonon 
interactions in silicene, moving beyond the limitations of harmonic or 
quasi-harmonic approximations for lattice vibrational modes. 
Such interactions can alter the electronic structure of 2D materials, 
in a manner similar to what has been observed in crystalline solids 
like diamond or cubic SiC \cite{ra06,ra08}. In these materials, 
nuclear quantum motion and anharmonic effects, particularly at 
low temperatures, lead to a renormalization (reduction) of the 
electronic band gap. For example, the direct gap at the $\Gamma$ point 
decreases by approximately 10\% as $T \to 0$.
In silicene, the electronic gap can be modulated through various methods, 
including the application of perpendicular electric fields, chemical 
functionalization (e.g., hydrogenation, oxidation), metal atom adsorption, 
or by using confined structures \cite{si-kh20}. A precise understanding of 
the electronic structure and its modifications due to electron-phonon 
interactions is critical for potential applications of silicene 
in electronic devices.
As in three-dimensional (3D) materials, the actual electronic gap in 
buckled silicene is expected to be influenced by nuclear quantum motion. 
This can result in changes of up to 10\% in the gap, a significant 
shift that would notably affect the electrical conductivity and other 
electronic properties. Further exploration of this 
subject is needed, both experimentally and theoretically, particularly 
through future studies utilizing PIMD simulations.

\section{Structural properties}

\subsection{Interatomic distance}

In the pursuit of the minimum-energy configuration of silicene at 
the classical zero temperature limit, the TB Hamiltonian reveals 
a distinctive buckled chair-like structure, characterized by an 
interatomic distance of $d_{\rm Si-Si} = 2.284$ \AA. 
The vertical displacement ($z$ coordinate) between adjacent 
silicon atoms, amounts to $h =$ 0.574~\AA. This structural 
feature emerges from the interplay between sp$^2$ and sp$^3$ hybridization, 
the latter being characteristic of bulk silicon with tetrahedral 
coordination of Si atoms.
Several DFT calculations corroborate 
the non-planar nature of silicene, predicting an out-of-plane buckling 
distance $h$ ranging between 0.44 and 0.49 \AA\ 
\cite{si-ca09,si-sa09,si-ga11,si-ka14,si-sc13}, along with a separation 
between nearest neighbors, $d_{\rm Si-Si}$, varying from 2.24 to 2.28 \AA\ 
\cite{si-ca09,si-sa09,si-ga11,si-sc13}.

\begin{figure}
\vspace{-7mm}
\includegraphics[width=7cm]{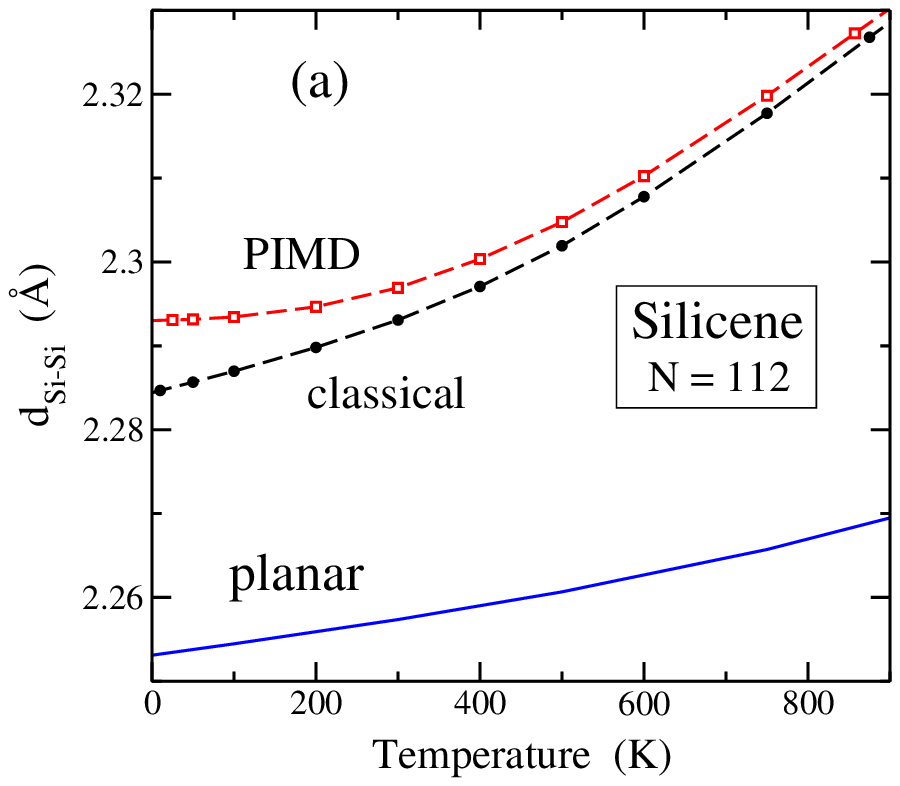}
\includegraphics[width=7cm]{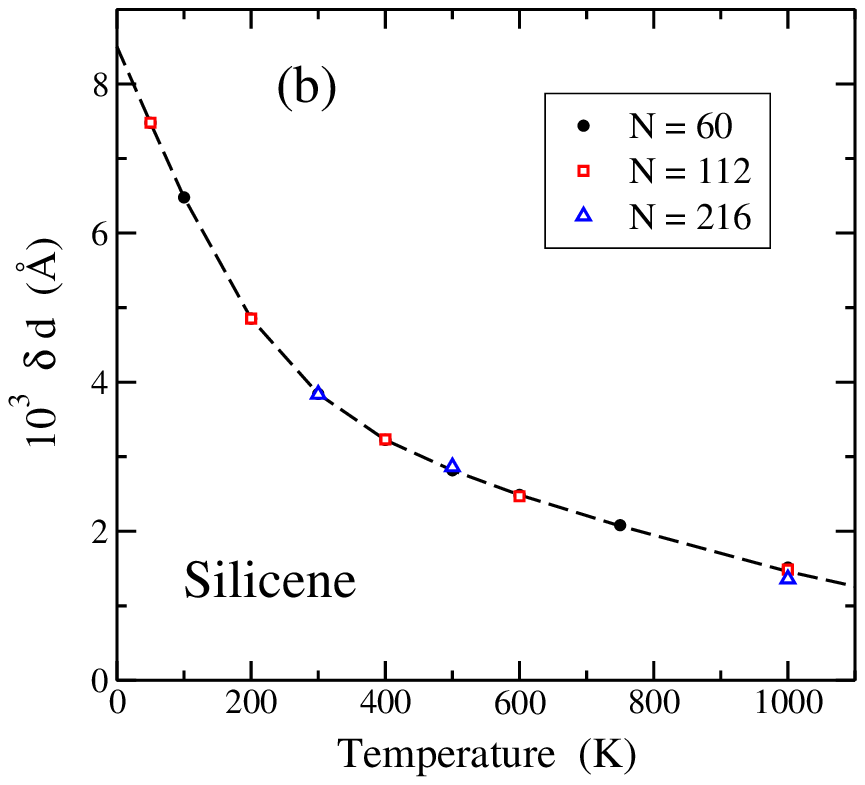}
\vspace{-5mm}
\caption{(a) Temperature dependence of the mean distance between
nearest-neighbor atoms in silicene, derived from classical
(solid circles) and PIMD simulations (open squares)
for $N = 112$. Error bars are smaller than the symbol size.
The solid line represents the interatomic distance obtained from
classical MD simulations of planar silicene.
(b) Difference $\delta d$ between quantum and classical results
for $N = 60$ (circles), 112 (squares), and 216 (triangles).
Dashed lines in (a) and (b) are guides to the eye.
}
\label{f6}
\end{figure}

In the case of planar silicene (buckling distance $h = 0$), our investigation 
reveals a minimum-energy configuration with an interatomic distance of 
$d_{\rm Si-Si}$ = 2.253~\AA, marginally shorter than the buckled counterpart. 
This flat configuration exhibits an energy 86~meV/atom higher than 
the absolute minimum, $E_0$. Such a difference forms a substantial 
energy barrier, effectively favoring the buckled structure.
It is noteworthy that at room temperature ($T$ = 300~K), the thermal 
energy per degree of freedom (e.g., along the $z$ coordinate) 
amounts to $k_B T =$ 26~meV/atom. 
This comparison underscores the energetic preference for buckled 
silicene, emphasizing the persistence of its structural stability even under 
thermal agitation.

Turning to the outcomes of our simulations at finite 
temperatures, Fig.~6(a) illustrates the distance $d_{\rm Si-Si}$ 
across the temperature range from $T = 0$ to 900~K. 
The data points, represented by solid circles and open squares, 
correspond to classical MD and PIMD simulations, respectively, 
both conducted with a system size of $N = 112$. 
Remarkably, the influence stemming from the finite cell size remains 
small compared to the statistical error bars, 
a consistent observation across both classical and quantum results 
(further discussed below).

In the classical approach, our findings confirm the expected linear 
relationship that characterizes bond expansion in crystalline solids 
at low temperatures \cite{ki05}. This linear trend holds up to approximately 
$T = 200$~K, with our results yielding a slope of 
$\partial d_{\rm Si-Si} / \partial T = 2.7 \times 10^{-5}$ \AA/K within 
this temperature range. However, beyond this threshold, significant 
deviations from linearity become apparent.
For example, at $T = 875$~K, the extrapolated Si-Si bond distance from 
the low-temperature linear fit, $d_{\rm Si-Si} = 2.308$\AA, contrasts with 
the actual value from MD simulations, $d_{\rm Si-Si} = 2.327$~\AA, indicating 
a relative increase of 0.8\%. This nonlinearity is more pronounced compared 
to similar materials such as 2D SiC \cite{he22} or graphene \cite{he16}, 
largely due to silicene's distinct buckled structure, which amplifies 
bond dilation with rising temperatures.
At 875~K, classical simulations for graphene and SiC monolayers show 
a relative difference of less than 0.1\% between the simulation results at 
this temperature and the linear extrapolation of the interatomic distance 
from low-temperature data. This underscores the significantly enhanced 
anharmonicity observed in buckled silicene compared to planar materials.

To obtain deeper insight into this departure from linearity, we carried out
classical simulations for planar silicene at various temperatures, while 
maintaining the $z$ coordinate of silicon atoms fixed at $z = 0$. 
The outcomes of these simulations, portraying the interatomic distance 
as a function of $T$, are depicted in Fig. 6(a) by a solid line. 
For $T \to 0$, $d_{\rm Si-Si}$ converges to a value of 2.253~\AA, aligning 
with the previously mentioned minimum-energy configuration for planar 
silicene. This value is smaller than that observed in unrestricted 
simulations, where atomic motion is free in the out-of-plane direction.
Earlier DFT calculations also corroborate a reduction in the interatomic 
distance for the planar configuration compared to the buckled minimum-energy 
structure of silicene. Indeed, reported values of $d_{\rm Si-Si}$ ranging 
between 2.22 and 2.25~\AA\ have been documented in such 
calculations \cite{si-ga11,si-sc13}.

Our simulations of planar silicene reveal a noteworthy trend in the relationship 
between $d_{\rm Si-Si}$ and $T$, appearing to approximate 
linearity across a broader temperature range compared to the buckled 
material, as illustrated in Fig. 6(a). Specifically, for planar silicene, 
we observe a slope of 
$\partial d_{\rm Si-Si} / \partial T = 1.4 \times 10^{-5}$ \AA/K at low 
temperatures. This value is approximately half of that 
obtained for the buckled counterpart (as discussed earlier).

The interplay between the anharmonicity inherent in the interatomic potential 
derived from the TB Hamiltonian and the atomic quantum dynamics is 
anticipated to induce an expansion in the equilibrium distance $d_{\rm Si-Si}$. 
This effect becomes evident through the noticeable gap observed between 
the curves representing PIMD and classical MD simulations in Fig.~6(a).
As $T \to 0$, extrapolations 
from PIMD simulations conducted at finite temperatures indicate a Si--Si 
distance of 2.293~\AA, indicative of a bond expansion 
$\Lambda_0 = 8.5 \times 10^{-3}$~\AA, attributed to zero-point motion.
The difference between quantum and classical data gradually diminishes 
with increasing temperature.
This reduction in difference can be attributed to the diminishing 
influence of quantum fluctuations as temperature rises, culminating 
in the convergence of the two curves at elevated temperatures.

The zero-point expansion $\Lambda_0$ found between low-temperature PIMD 
and classical MD simulation data, amounting to 0.4\% of the bond 
length, can be compared to relative quantum-to-classical expansions 
reported from similar TB calculations for related materials. Specifically, 
these expansions are 0.5\% in graphene \cite{he20b} and 0.3\% in 2D 
silicon carbide \cite{he22}. The relative values of $\Lambda_0$ for 
the three 2D materials are of the same order of magnitude, despite 
the large variation in interatomic distances for C--C, C--Si, and 
Si--Si bonds (1.429, 1.720, and 2.284 \AA, respectively). Notably, 
this relative quantity $\Lambda_0$ reaches its minimum value for 
the binary compound SiC.

In Fig.~6(b), we illustrate the temperature-dependent behavior of 
the interatomic bond expansion, $\delta d$, resulting from quantum motion. 
This expansion is derived from the difference between the outcomes of PIMD 
and classical MD simulations at each temperature.
Starting from the zero-point value $\Lambda_0$, $\delta d$ gradually diminishes 
to $1.5 \times 10^{-3}$ \AA\ at 1000~K, indicating a reduction by a factor 
of approximately 6 within this temperature range. 
At $T = 300$~K, $\delta d$ measures $3.8 \times 10^{-3}$ \AA, 
equivalent to 45\% of $\Lambda_0$.
The plot presents data for system sizes $N = 60$ (depicted by solid circles), 
112 (open squares), and 216 (open triangles). 
For each temperature, the data obtained for different cell sizes 
coincide within the error bars, and in most instances 
they appear superimposed in Fig.~6(b).

When examining the buckling distance $h$, our PIMD simulations reveal 
a measurement of 0.579~\AA\ at low temperatures, slightly higher than 
the classical counterpart of 0.574~\AA. This increase in $h$ equates 
to roughly 1\% of the classical value. However, with the escalation 
of temperature, we observe a reduction in the buckling distance, 
yielding $h =$~0.557 and 0.559~\AA\ at $T = 1000$~K from classical and 
quantum simulations, respectively.
It is important to emphasize that these data reflect average values 
computed over long simulation runs. The actual difference in 
$z$ coordinates between silicon atoms fluctuates notably within each 
configuration, as does the mean value across various simulation steps, 
particularly at elevated temperatures. This variability primarily stems 
from the bending of the silicene layer under thermal agitation.

\begin{figure}
\vspace{-7mm}
\includegraphics[width=7cm]{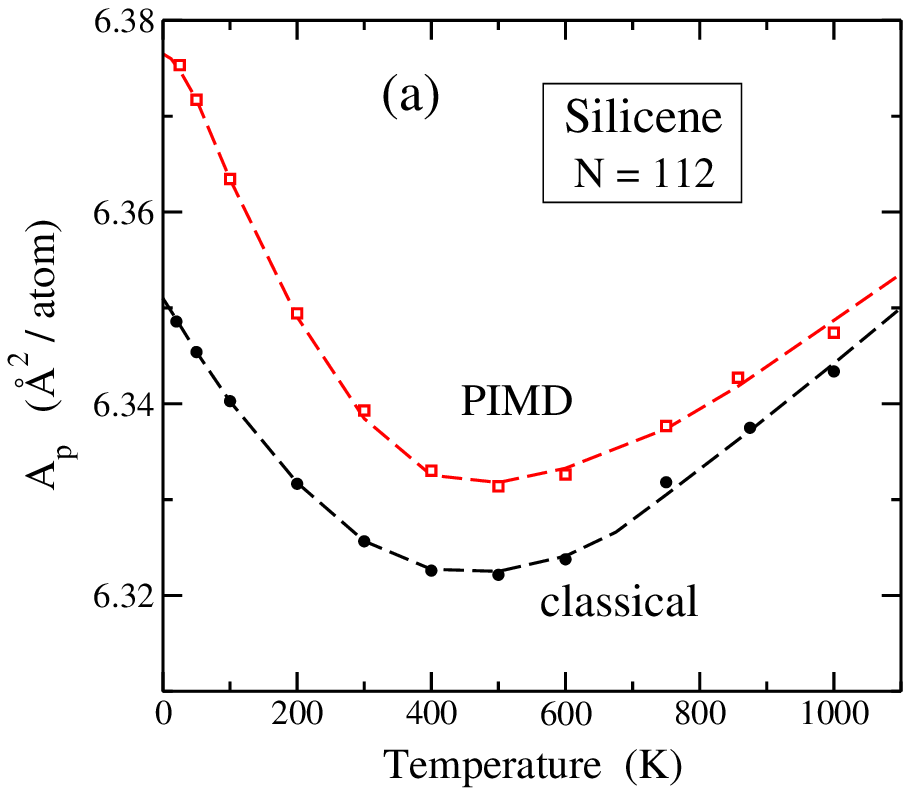}
\includegraphics[width=7cm]{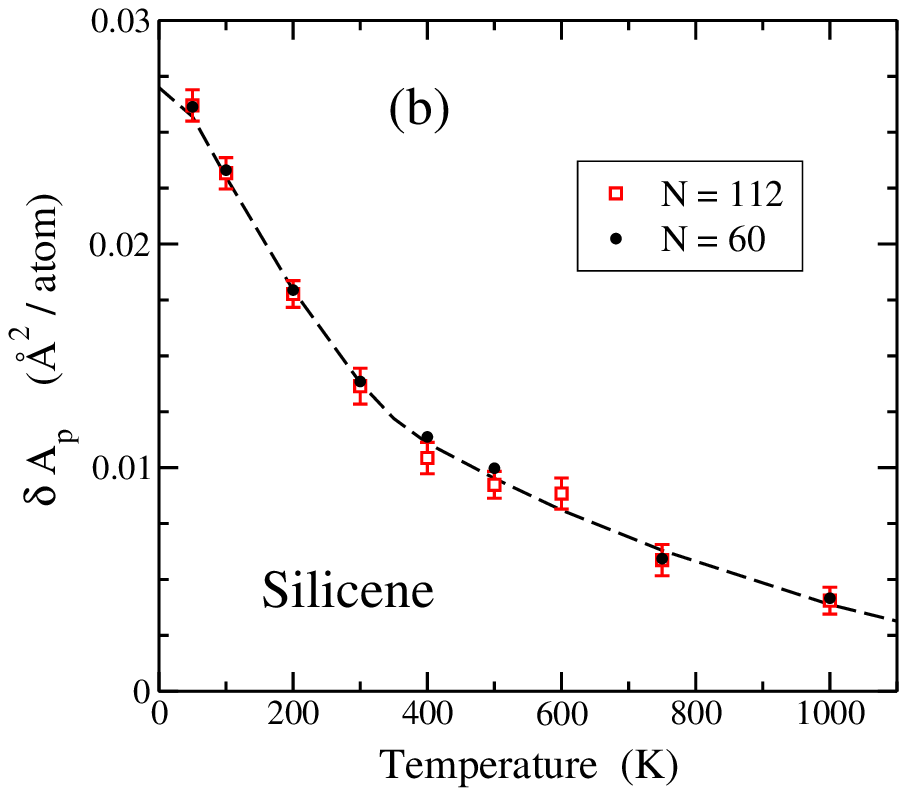}
\vspace{-5mm}
\caption{(a) In-plane area, $A_p$, as a function of temperature, derived
from classical (circles) and PIMD simulations (open squares).
Error bars are in the order of the symbol size.
(b) Difference, $\delta A_p$, between quantum and classical results
vs $T$ for $N = 60$ (solid circles) and 112 (open squares).
Lines are guides to the eye.
}
\label{f7}
\end{figure}

\subsection{In-plane area}

Simulations performed in the isothermal-isobaric ensemble afford 
three-dimensional freedom to the Si atoms to move in space. 
This setup allows us to calculate 
the in-plane area $A_p$ of silicene and its behavior with respect 
to temperature.  Note that $A_p$ is the conjugate variable to the 
in-plane stress $\tau$, which we impose to vanish in our simulations.

In Fig.~7(a), we illustrate the in-plane area $A_p$ as a function of 
temperature, derived from classical MD (circles) and PIMD simulations 
(open squares). At low temperatures, the classical results converge 
to the minimum-energy state, characterized by an in-plane area 
$A_p^0$ = 6.351 \AA$^2$/atom.
The length $d_{xy}$ of a Si--Si bond projection on the $xy$ plane satisfies 
$d_{xy}^2 = d_{\rm Si-Si}^2 - h^2$, and 
$A_p = 3 \sqrt{3} \, d_{xy}^2 / 4$.
Across both types of simulations, the in-plane area $A_p$ decreases
in the temperature range from $T = 0$ to 400~K, 
indicating a contraction with  $\partial A_p / \partial T < 0$.
In the classical data, a minimum in $A_p$ is evident at a temperature 
$T_m \approx 450$~K, whereas the quantum result exhibits a slightly higher 
value around 500~K. This trend of $T_m$ increasing when nuclear quantum 
effects are considered has been previously observed in graphene \cite{he16}.

The behavior of the in-plane area $A_p$ with temperature is governed 
by the interplay of two competing factors. Firstly, the atomic motion 
out of the plane induces a reduction in $A_p$, a phenomenon 
linked to the mean-square fluctuation (MSF), $(\Delta z)^2$, at each 
temperature \cite{he23}. Secondly, thermal expansion of the Si--Si bonds 
results in an enlargement of the crystalline membrane, promoting an increase 
in the actual in-plane area. At low temperatures, the former effect 
predominates, leading to a negative rate of change: 
$\partial A_p / \partial T < 0$.
While the MSF $(\Delta z)^2$ does increase with temperature, this increment 
follows a sublinear trend \cite{ga14,he23}. In contrast, the elongation 
of the Si--Si bond length with rising temperature exhibits a superlinear 
trend (see Fig.~6(a)). Consequently, as temperature increases, 
the dominance shifts towards the second mechanism, resulting in 
$\partial A_p / \partial T > 0$.

Furthermore, in Fig.~7(b), we depict the variance between quantum and 
classical outcomes for $N = 60$ and 112 atoms. The figure notably 
illustrates the convergence of classical results towards quantum outcomes 
as temperature rises. The disparity in area between quantum and classical 
findings diminishes from its zero-point value, 
$\delta A_p^0 = 2.7 \times 10^{-2}$~\AA$^2$/atom, with increasing 
temperature.
At $T = 1000$~K, this difference narrows to $4 \times 10^{-3}$ \AA$^2$/atom, 
representing a sevenfold reduction in quantum expansion compared to 
the zero-temperature limit. This consistent trend is observable across 
different sizes of the simulation cell, as illustrated in Fig.~7(b).

For planar silicene, the area $A_p$ is determined to be 6.595~\AA$^2$/atom 
for the minimum-energy configuration (classical, $T = 0$). 
In this scenario a thermal expansion of the 2D material 
becomes apparent when it is strictly confined to the $z = 0$ plane, 
contrasting with the reduction of $A_p$ observed for buckled silicene 
at low temperatures. This difference arises as a direct consequence 
of the inhibition of bending in the planar configuration.
At low temperatures, we observe a positive slope of 
$\partial A_p / \partial T = 5 \times 10^{-5}$ \AA$^2$/K, a significant 
departure from the negative slope of $-1.2 \times 10^{-4}$ \AA$^2$/K 
corresponding to classical results for buckled silicene.

We conclude this section by emphasizing the importance of in-plane area 
variations in ensuring compatibility with other materials in epitaxial 
structures \cite{si-mo18}. Specifically, the increase in $A_p$ due to 
nuclear quantum motion can impact the functionality of potential devices. 
Lattice mismatches between adjacent material layers may lead to the 
formation of misfit dislocations, which can degrade microelectronic 
device performance by introducing undesirable electrical properties or 
compromising mechanical stability.
Thus, understanding these area changes is crucial for optimizing the 
functionality of epitaxial structures in practical applications. 
In the case of silicene, the quantum effects related to in-plane area 
(or lattice parameters) discussed here can significantly influence 
its growth and structural properties in epitaxial systems. 
These effects could be explored on various substrates, such as silver, 
aluminum, or MoS$_2$ \cite{si-ma23}, where small lattice mismatches 
(on the order of 1\% or even less) may play a crucial role in 
determining the stability and performance of silicene-based structures.

\begin{figure}
\vspace{-7mm}
\includegraphics[width=7cm]{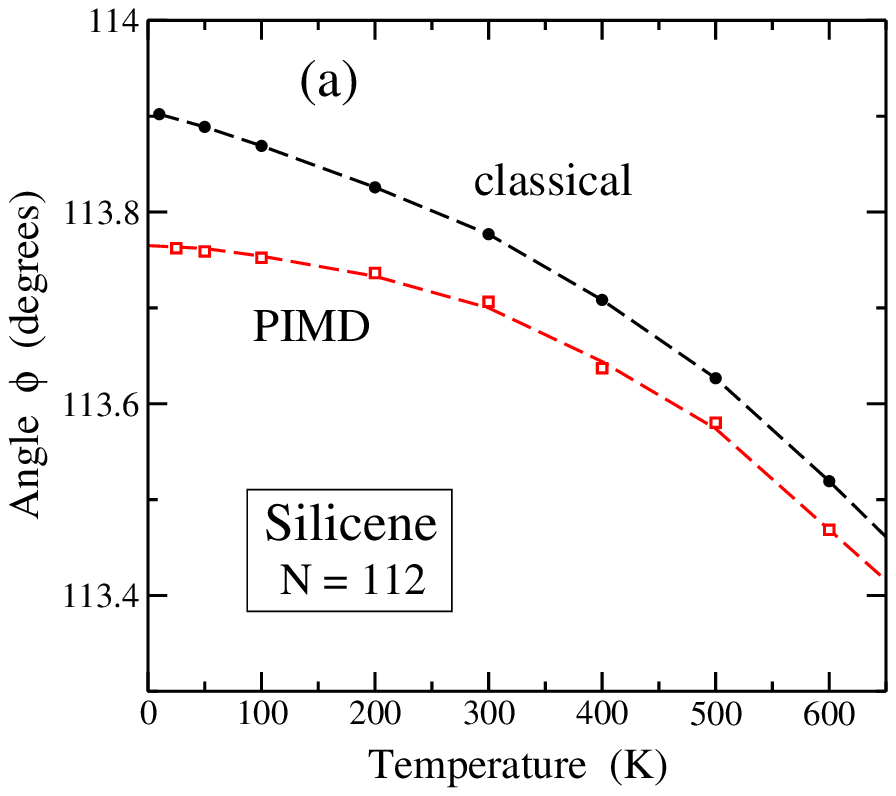}
\includegraphics[width=7cm]{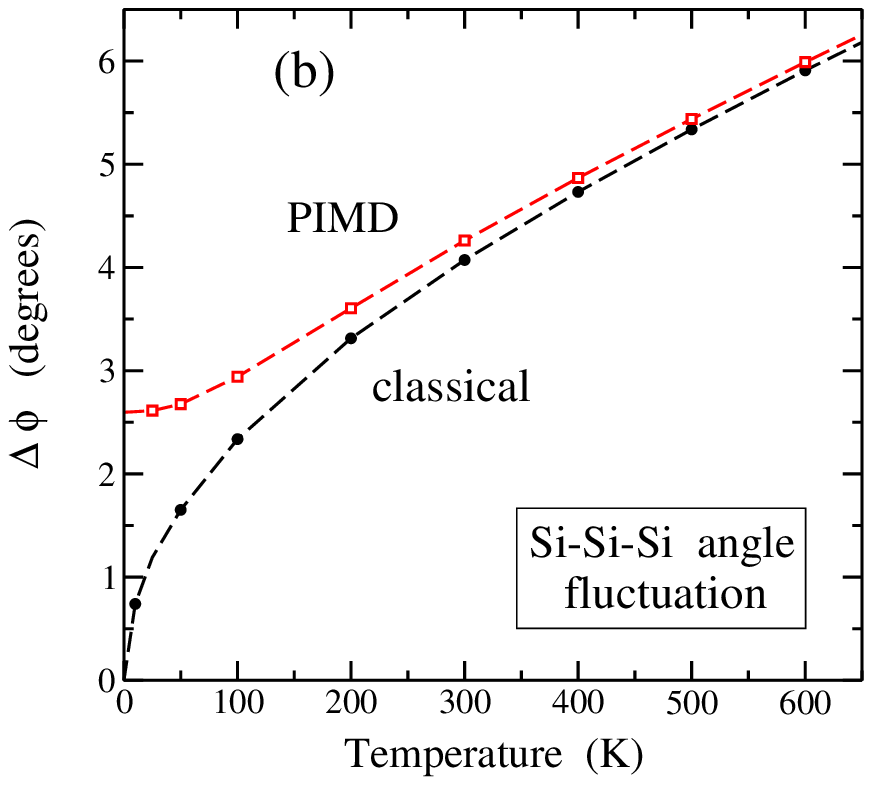}
\vspace{-5mm}
\caption{(a) Temperature dependence of (a) the mean Si-Si-Si angle
$\phi$ in silicene, and (b) its root-mean-square fluctuation
$\Delta \phi$.
Solid circles and open squares represent results from classical MD
and PIMD simulations, respectively.
Dashed lines are guides to the eye.
Error bars are on the order of the symbol size.
}
\label{f8}
\end{figure}

\subsection{Si-Si-Si angle}

In bulk silicon, which adopts a face-centered cubic diamond structure, 
the bond angle between an atom and two of its nearest neighbors is 
$\phi = 109.47^{\circ}$, indicative of $sp^3$ electronic hybridization 
in Si atoms. Conversely, in a planar 2D atomic layer with 
a honeycomb structure, such as graphene or strictly flat silicene, 
the bond angle $\phi$ is $120^{\circ}$, reflecting $sp^2$ hybridization.
In the case of silicene, which exhibits an actual buckled 
structure, the bond angle $\phi$ assumes intermediate values between 
these two limits due to the interplay between $sp^2$ and $sp^3$ 
hybridization in this material.

The Si-Si-Si angle in silicene is subject to fluctuations due to 
the combined effects of thermal and quantum motion affecting the silicon 
atoms. In Fig.~8(a), we illustrate the temperature-dependent variation of 
the average angle $\phi$ as determined from classical MD simulations 
(depicted as solid circles) and PIMD simulations (open squares).
Across both simulation methods, one observes that the angle decreases 
with increasing temperature, exhibiting a tendency to converge towards 
each other at elevated temperatures. In the $T = 0$ classical limit, 
we find $\phi = 113.90^{\circ}$, which notably aligns closer 
to the angle corresponding to tetrahedral 3D coordination rather than 
the hexagonal 2D case (the mean value for these two configurations 
being $114.7^{\circ}$).
Upon extrapolating the PIMD simulation results towards zero 
temperature ($T \to 0$), the angle $\phi$ is estimated to be 
$\phi = 113.76^{\circ}$, somewhat smaller than the classical minimum.

From a purely geometric perspective, the angle $\phi$ can be directly 
correlated with both the interatomic bond length $d_{\rm Si-Si}$ and 
the buckling distance $h$. Specifically, in the absence of atomic motion, 
as in the classical $T = 0$ limit, this relationship is succinctly 
expressed by:
\begin{equation}
    \cos \phi = \frac12 (3 c^2 - 1),
\label{angphi}
\end{equation}
where $c = h / d_{\rm Si-Si}$. It is noteworthy that Eq.~(\ref{angphi}) 
reproduces the expected angles for distinct structural 
configurations: for a planar sheet ($c = 0$), it yields $\phi = 120^{\circ}$, 
while for the tetrahedral 3D arrangement ($c = 1/3$), it predicts 
$\cos \phi = -1/3$.
However, it is important to acknowledge that in the presence of atomic motion, 
whether classical thermal or quantum, the accuracy of Eq.~(\ref{angphi}) 
diminishes. This is primarily attributed to the induced 
bending within the silicene layer.

In Fig.~8(b), we illustrate the root-mean-square fluctuation $\Delta \phi$ 
of the Si-Si-Si angle, as determined from classical MD 
(depicted as solid circles) and PIMD simulations (open squares).
For $T \to 0$, $\Delta \phi$ converges to zero for the classical data.
Conversely, for the quantum results, 
a value of 2.6$^{\circ}$ is observed due to zero-point motion.
With increasing temperature, $\Delta \phi$ escalates rapidly. At $T = 1000$~K 
(not shown in the figure), it reaches approximately 8$^{\circ}$ for both 
classical and quantum data. At this temperature, the difference
between the two is less than 0.05$^{\circ}$.

\section{Compressibility}

Mechanical and elastic properties are crucial for understanding the strength,
reliability, and design of silicene-based devices. A detailed understanding
of the mechanical response provides valuable insights for applications in
various systems, including those used in nanoelectronics
\cite{si-mo17,si-zh12,si-yo21,si-ze18}. Additionally, structural
changes or deformations can impact the performance of these applications
by altering the electronic band structure.

PIMD simulations offer valuable insights into the elastic 
characteristics of solids, particularly in the realm of 2D crystalline 
membranes. These properties undergo significant influence from quantum 
nuclear motion, particularly noticeable at temperatures below 
the Debye temperature of the materials under investigation.

In this section, we examine the in-plane compressibility of silicene, 
or its inverse, the 2D modulus of hydrostatic compression $B_p$. 
For layered materials, the isothermal $B_p$ 
at temperature $T$ is defined as \cite{be96b}:
\begin{equation}
  B_p = - A_p \left( \frac{\partial \tau}{\partial A_p} \right)_T   \, .
\label{bulk1}
\end{equation}
Here, the in-plane biaxial pressure $\tau$ and area $A_p$ are variables 
associated with the layer plane, and in the isothermal-isobaric 
ensemble they behave as conjugate variables. 
It is worth noting that $B_p$ is linked to the elastic stiffness 
constants of silicene, as expressed by: 
$B_p = (C_{11} + C_{12}) / 2$ \cite{be96b,he23}.

The modulus $B_p$ can also be determined using the fluctuation formula 
\cite{la80,he18b}:
\begin{equation}
    B_p = \frac{k_B T A_p}{N (\Delta A_p)^2}   \; ,
\label{bulk2}
\end{equation}
where $(\Delta A_p)^2$ represents the MSF of 
the area $A_p$, calculated here from classical and quantum simulations 
under $\tau = 0$. This expression offers an alternative, feasible approach 
to compute $B_p$, circumventing the need for deriving 
$(\partial A_p / \partial \tau)_T$ through numerical methods, which 
necessitate additional simulations at pressures near $\tau = 0$.
We have verified at several temperatures that both methods produce $B_p$ 
results consistent within the statistical error bars of the simulations. 
This validation serves as a valuable consistency check for the numerical 
techniques employed in this study.

\begin{figure}
\vspace{-7mm}
\includegraphics[width=7cm]{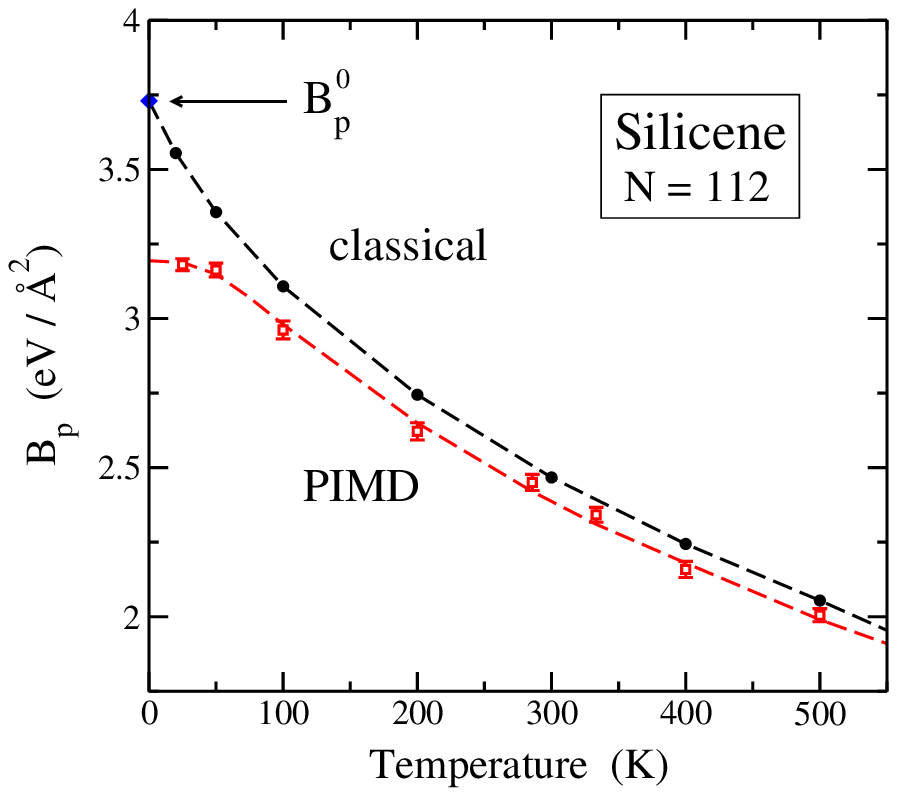}
\vspace{-5mm}
\caption{Two-dimensional modulus of compression $B_p$ of silicene,
found from PIMD (open squares) and classical MD simulations
(solid circles) for $N = 112$. Error bars of the classical data
are on the order of the symbol size.
A diamond at $T = 0$, indicated by an arrow, shows the $B_p^0$ value
obtained from Eq.~(\ref{bulk0}) for the minimum-energy configuration.
Dashed lines are guides to the eye.
}
\label{f9}
\end{figure}

In Fig. 9, we present the modulus $B_p$ of silicene as a function of 
temperature, derived from our PIMD simulations (open squares) and classical 
MD (solid circles), using Eq.~(\ref{bulk2}).
The classical 2D modulus $B_p$, as obtained from our simulations, 
demonstrates a notable decrease with increasing temperature. 
Specifically, at low temperatures, we observe a slope of 
$\partial B_p / \partial T = -7.5 \times 10^{-3}$ eV/(\AA$^2$~K). 
However, this slope decreases as the temperature rises, reaching 
$\partial B_p / \partial T = -1.9 \times 10^{-3}$ eV/(\AA$^2$~K) at 500~K.

In a classical 2D solid, $B_p$ at $T = 0$ is determined by 
the expression:
\begin{equation}
 B_p^0 =  A_p \, \left. \frac{\partial^2 E}{\partial A_p^2}
           \right|_{A_p^0}  \; ,
\label{bulk0}
\end{equation}
where $E$ represents the energy. For silicene, $B_p^0$ is calculated to be 
3.73~eV/\AA$^2$, consistent with the extrapolation for $T \to 0$ of 
the $B_p$ data derived from our classical simulations. This low-temperature 
limit of the compression modulus is indicated by an arrow in Fig.~9.
As temperature rises, our simulations show decreasing $B_p$ values, 
as previously noted. For instance, at $T = 750$ K, $B_p$ is determined 
to be 1.65~eV/\AA$^2$, which represents 44\% of the zero-temperature 
limit, $B_p^0$.

For planar silicene, employing Eq.~(\ref{bulk0}), we derive 
$B_p^0 =$ 6.2 eV/\AA$^2$, significantly greater than the equivalent 
value for the buckled material. This disparity arises because Si--Si 
bonds in the latter are not aligned parallel to the layer plane, rendering 
the actual material softer than the ideal planar configuration.

In our quantum results for buckled silicene, shown in Fig.~9 (open squares), 
we notice a significant decrease in $B_p$ at low temperatures compared 
to the classical data. Extrapolating the quantum data to $T = 0$ 
yields $B_p = 3.19$ eV/\AA$^2$, contrasting with the classical 
low-$T$ value: $B_p^0 = 3.73$ eV/\AA$^2$. This implies a reduction of 
14\% due to zero-point quantum motion.
The disparity between classical and quantum outcomes diminishes with 
increasing temperature, and at $T = 500$~K it is about 2\%.

We highlight that the derivative $\partial B_p / \partial T$ 
must vanish in the limit $T \to 0$, in accordance with the third law 
of Thermodynamics \cite{ca85,be00b}. While this condition is met 
by PIMD results, classical low-temperature data exhibit a negative slope, 
violating this principle.

In comparison to the results for silicene, it is worth noting that 
similar calculations performed for monolayer graphene and 2D SiC yield 
$B_p^0$ values of 12.7 and 5.5 eV/\AA$^2$, respectively \cite{he22}. 
These values, particularly that of graphene, are significantly higher 
than the corresponding result for silicene. Specifically, the $B_p^0$ 
value for graphene is more than three times larger than that of silicene. 
A substantial portion of this difference arises from the disparity in 
bond strength, with C--C bonds being considerably stronger than Si--Si bonds. 
Furthermore, $B_p$ in silicene is further diminished by its nonplanar 
structure, as previously discussed, with the buckled configuration 
resulting in a reduction of about 40\% compared to the planar form.

In this study, we have shown that nuclear quantum motion in silicene leads 
to a significant reduction in the 2D compression modulus, particularly 
at low temperatures, where a decrease of 14\% is observed. This reduction 
can impact the stability of silicene layers in epitaxial structures, 
where tensile or compressive stresses may arise due to lattice mismatches 
between different components. At higher temperatures, the average Si-Si 
bond distance, $d_{\rm Si-Si}$, increases (see Fig.~6), weakening the bonds 
and making them more susceptible to breaking under lower levels of tensile 
strain. Additionally, the Si-Si bond distances exhibit larger fluctuations 
at elevated temperatures, further contributing to bond breakage.

To conclude this section, it is important to highlight that, although 
DFT and other computational methods provide reasonable predictions for 
mechanical properties like the modulus $B_p$, they usually do not consider 
the quantum dynamics of atomic nuclei. When this omission is combined with 
anharmonicity of lattice vibrational modes, significant changes in 
these properties can arise. This phenomenon affects the modulus $B_p$, 
as discussed above, and is also expected to influence the in-plane Young's 
modulus $E_p$. In fact, the latter is related to $B_p$ by the relation 
$E_p = 2 B_p (1 - \nu)$, where $\nu$ represents the Poisson's ratio. 
Both $B_p$ and $E_p$ are experimentally measurable quantities, and 
they exhibit nuclear quantum effects, particularly at low temperatures.
Comparing computational results with experimental measurements on 
silicene samples poses a challenge for future research, as 
significant advancements have been made in recent years, especially in 
studies involving 2D materials under various conditions. For instance, 
the mechanical properties of graphene have been investigated using 
phonon dispersion and low-energy electron diffraction (LEED) techniques 
on both free-standing and supported samples \cite{po12,po15}.
Moreover, it is noteworthy that in the case of silicene, the relative 
reduction in the modulus $B_p$ due to nuclear quantum motion at low 
temperatures (14\%) is slightly larger than that observed for graphene 
(12\%) in similar PIMD simulations, despite the fact that $B_p^0$ for 
silicene is significantly lower than that of graphene. This greater 
reduction in silicene may enhance the experimental detection of such 
quantum effects, particularly in the low-temperature regime.  \\

\section{Summary}

In this paper, we  have shown
that PIMD simulations provide a powerful means to assess the impact 
of nuclear quantum motion on diverse structural and vibrational features 
of 2D crystalline solids. In the case of silicene, with a Debye 
temperature around 600~K, we have observed significant nuclear quantum 
effects even for $T$ exceeding room temperature.

At low temperatures, silicene's behavior is influenced by 
a combination of anharmonicity and quantum motion. 
We have investigated this behavior using a tight-binding Hamiltonian. 
To gauge the extent of quantum effects, we have compared data 
from classical MD and PIMD simulations. Additionally, a harmonic 
approximation serves as a reference point for assessing 
the magnitude of anharmonicity in various physical properties 
of silicene.

The quantization of lattice vibrations in silicene induces 
alterations in the average interatomic distances and in-plane area. 
The zero-point expansion of $A_p$ corresponds to a 0.4\% deviation 
from the classical calculation, a relative value akin to the bond 
dilation observed for $T \to 0$.

The quantum correction to the 2D compression modulus $B_p$ at low 
temperatures is calculated to be $-0.54$ eV/\AA$^2$, equivalent to 
a 14\% reduction from the classical value $B_p^0$. 
This decrease in $B_p$ resulting from nuclear quantum motion 
underscores the softer nature 
of real silicene compared to classical predictions, particularly 
evident at low $T$.

We highlight the alignment of low-temperature PIMD simulation results 
with the third law of Thermodynamics. This entails that the temperature 
derivative of various magnitudes (such as $d_{\rm S-Si}$, $A_p$, $B_p$) 
must approach zero as temperature tends to zero. 
Notably, this feature is absent in classical simulations, 
a distinction that has not received ample emphasis in the literature
about atomistic simulations of materials.

PIMD simulations, similar to those described in this study, have 
the potential to provide valuable insights into the interplay between 
anharmonicity and quantization of vibrational modes in stressed 
silicene. Additionally, they can elucidate similar dynamics in silicane, 
a hydrogenated silicene layer.

\begin{acknowledgments}
We express our gratitude to Rafael Ram\'irez for his invaluable discussions 
and assistance with computational methods.
This work was supported by Ministerio de Ciencia e Innovaci\'on 
(Spain) under Grant Number PID2022-139776NB-C66.
\end{acknowledgments}



\begin{thebibliography}{98}
\expandafter\ifx\csname natexlab\endcsname\relax\def\natexlab#1{#1}\fi
\expandafter\ifx\csname bibnamefont\endcsname\relax
  \def\bibnamefont#1{#1}\fi
\expandafter\ifx\csname bibfnamefont\endcsname\relax
  \def\bibfnamefont#1{#1}\fi
\expandafter\ifx\csname citenamefont\endcsname\relax
  \def\citenamefont#1{#1}\fi
\expandafter\ifx\csname url\endcsname\relax
  \def\url#1{\texttt{#1}}\fi
\expandafter\ifx\csname urlprefix\endcsname\relax\def\urlprefix{URL }\fi
\providecommand{\bibinfo}[2]{#2}
\providecommand{\eprint}[2][]{\url{#2}}

\bibitem[{\citenamefont{Zandvliet}(2014)}]{si-za14}
\bibinfo{author}{\bibfnamefont{H.~J.~W.} \bibnamefont{Zandvliet}},
  \bibinfo{journal}{Nano Today} \textbf{\bibinfo{volume}{9}},
  \bibinfo{pages}{691} (\bibinfo{year}{2014}).

\bibitem[{\citenamefont{Tao et~al.}(2015)\citenamefont{Tao, Cinquanta, Chiappe,
  Grazianetti, Fanciulli, Dubey, Molle, and Akinwande}}]{si-ta15}
\bibinfo{author}{\bibfnamefont{L.}~\bibnamefont{Tao}},
  \bibinfo{author}{\bibfnamefont{E.}~\bibnamefont{Cinquanta}},
  \bibinfo{author}{\bibfnamefont{D.}~\bibnamefont{Chiappe}},
  \bibinfo{author}{\bibfnamefont{C.}~\bibnamefont{Grazianetti}},
  \bibinfo{author}{\bibfnamefont{M.}~\bibnamefont{Fanciulli}},
  \bibinfo{author}{\bibfnamefont{M.}~\bibnamefont{Dubey}},
  \bibinfo{author}{\bibfnamefont{A.}~\bibnamefont{Molle}}, \bibnamefont{and}
  \bibinfo{author}{\bibfnamefont{D.}~\bibnamefont{Akinwande}},
  \bibinfo{journal}{Nature Nanotech.} \textbf{\bibinfo{volume}{10}},
  \bibinfo{pages}{227} (\bibinfo{year}{2015}).

\bibitem[{\citenamefont{Grazianetti et~al.}(2016)\citenamefont{Grazianetti,
  Cinquanta, and Molle}}]{si-gr16}
\bibinfo{author}{\bibfnamefont{C.}~\bibnamefont{Grazianetti}},
  \bibinfo{author}{\bibfnamefont{E.}~\bibnamefont{Cinquanta}},
  \bibnamefont{and} \bibinfo{author}{\bibfnamefont{A.}~\bibnamefont{Molle}},
  \bibinfo{journal}{2D Mater.} \textbf{\bibinfo{volume}{3}},
  \bibinfo{pages}{012001} (\bibinfo{year}{2016}).

\bibitem[{\citenamefont{Molle et~al.}(2018)\citenamefont{Molle, Grazianetti,
  Tao, Taneja, Alam, and Akinwande}}]{si-mo18}
\bibinfo{author}{\bibfnamefont{A.}~\bibnamefont{Molle}},
  \bibinfo{author}{\bibfnamefont{C.}~\bibnamefont{Grazianetti}},
  \bibinfo{author}{\bibfnamefont{L.}~\bibnamefont{Tao}},
  \bibinfo{author}{\bibfnamefont{D.}~\bibnamefont{Taneja}},
  \bibinfo{author}{\bibfnamefont{M.~H.} \bibnamefont{Alam}}, \bibnamefont{and}
  \bibinfo{author}{\bibfnamefont{D.}~\bibnamefont{Akinwande}},
  \bibinfo{journal}{Chem. Soc. Rev.} \textbf{\bibinfo{volume}{47}},
  \bibinfo{pages}{6370} (\bibinfo{year}{2018}).

\bibitem[{\citenamefont{Zhuang et~al.}(2015)\citenamefont{Zhuang, Xu, Feng, Li,
  Wang, and Du}}]{si-zh15}
\bibinfo{author}{\bibfnamefont{J.}~\bibnamefont{Zhuang}},
  \bibinfo{author}{\bibfnamefont{X.}~\bibnamefont{Xu}},
  \bibinfo{author}{\bibfnamefont{H.}~\bibnamefont{Feng}},
  \bibinfo{author}{\bibfnamefont{Z.}~\bibnamefont{Li}},
  \bibinfo{author}{\bibfnamefont{X.}~\bibnamefont{Wang}}, \bibnamefont{and}
  \bibinfo{author}{\bibfnamefont{Y.}~\bibnamefont{Du}}, \bibinfo{journal}{Sci.
  Bulletin} \textbf{\bibinfo{volume}{60}}, \bibinfo{pages}{1551}
  (\bibinfo{year}{2015}).

\bibitem[{\citenamefont{Kharadi et~al.}(2020)\citenamefont{Kharadi, Malik,
  Khanday, Shah, Mittal, and Kaushik}}]{si-kh20}
\bibinfo{author}{\bibfnamefont{M.~A.} \bibnamefont{Kharadi}},
  \bibinfo{author}{\bibfnamefont{G.~F.~A.} \bibnamefont{Malik}},
  \bibinfo{author}{\bibfnamefont{F.~A.} \bibnamefont{Khanday}},
  \bibinfo{author}{\bibfnamefont{K.~A.} \bibnamefont{Shah}},
  \bibinfo{author}{\bibfnamefont{S.}~\bibnamefont{Mittal}}, \bibnamefont{and}
  \bibinfo{author}{\bibfnamefont{B.~K.} \bibnamefont{Kaushik}},
  \bibinfo{journal}{ECS J. Solid State Sci. Techn.}
  \textbf{\bibinfo{volume}{9}}, \bibinfo{pages}{115031} (\bibinfo{year}{2020}).

\bibitem[{\citenamefont{Tantardini et~al.}(2021)\citenamefont{Tantardini,
  Kvashnin, Gatti, Yakobson, and Gonze}}]{si-ta21}
\bibinfo{author}{\bibfnamefont{C.}~\bibnamefont{Tantardini}},
  \bibinfo{author}{\bibfnamefont{A.~G.} \bibnamefont{Kvashnin}},
  \bibinfo{author}{\bibfnamefont{C.}~\bibnamefont{Gatti}},
  \bibinfo{author}{\bibfnamefont{B.}~\bibnamefont{Yakobson}, \bibfnamefont{I}},
  \bibnamefont{and} \bibinfo{author}{\bibfnamefont{X.}~\bibnamefont{Gonze}},
  \bibinfo{journal}{ACS Nano} \textbf{\bibinfo{volume}{15}},
  \bibinfo{pages}{6861} (\bibinfo{year}{2021}).

\bibitem[{\citenamefont{Ghosal et~al.}(2023)\citenamefont{Ghosal,
  Bandyopadhyay, Chowdhury, and Jana}}]{si-gh23}
\bibinfo{author}{\bibfnamefont{S.}~\bibnamefont{Ghosal}},
  \bibinfo{author}{\bibfnamefont{A.}~\bibnamefont{Bandyopadhyay}},
  \bibinfo{author}{\bibfnamefont{S.}~\bibnamefont{Chowdhury}},
  \bibnamefont{and} \bibinfo{author}{\bibfnamefont{D.}~\bibnamefont{Jana}},
  \bibinfo{journal}{Rep. Progr. Phys.} \textbf{\bibinfo{volume}{86}},
  \bibinfo{pages}{096502} (\bibinfo{year}{2023}).

\bibitem[{\citenamefont{Ni et~al.}(2012)\citenamefont{Ni, Liu, Tang, Zheng,
  Zhou, Qin, Gao, Yu, and Lu}}]{si-ni12}
\bibinfo{author}{\bibfnamefont{Z.}~\bibnamefont{Ni}},
  \bibinfo{author}{\bibfnamefont{Q.}~\bibnamefont{Liu}},
  \bibinfo{author}{\bibfnamefont{K.}~\bibnamefont{Tang}},
  \bibinfo{author}{\bibfnamefont{J.}~\bibnamefont{Zheng}},
  \bibinfo{author}{\bibfnamefont{J.}~\bibnamefont{Zhou}},
  \bibinfo{author}{\bibfnamefont{R.}~\bibnamefont{Qin}},
  \bibinfo{author}{\bibfnamefont{Z.}~\bibnamefont{Gao}},
  \bibinfo{author}{\bibfnamefont{D.}~\bibnamefont{Yu}}, \bibnamefont{and}
  \bibinfo{author}{\bibfnamefont{J.}~\bibnamefont{Lu}}, \bibinfo{journal}{Nano
  Lett.} \textbf{\bibinfo{volume}{12}}, \bibinfo{pages}{113}
  (\bibinfo{year}{2012}).

\bibitem[{\citenamefont{Guo et~al.}(2015)\citenamefont{Guo, Zhang, Xiang, Gong,
  and Oshiyama}}]{si-gu15b}
\bibinfo{author}{\bibfnamefont{Z.-X.} \bibnamefont{Guo}},
  \bibinfo{author}{\bibfnamefont{Y.-Y.} \bibnamefont{Zhang}},
  \bibinfo{author}{\bibfnamefont{H.}~\bibnamefont{Xiang}},
  \bibinfo{author}{\bibfnamefont{X.-G.} \bibnamefont{Gong}}, \bibnamefont{and}
  \bibinfo{author}{\bibfnamefont{A.}~\bibnamefont{Oshiyama}},
  \bibinfo{journal}{Phys. Rev. B} \textbf{\bibinfo{volume}{92}},
  \bibinfo{pages}{201413} (\bibinfo{year}{2015}).

\bibitem[{\citenamefont{Zhu and Schwingenschlogl}(2016)}]{si-zh16}
\bibinfo{author}{\bibfnamefont{J.}~\bibnamefont{Zhu}} \bibnamefont{and}
  \bibinfo{author}{\bibfnamefont{U.}~\bibnamefont{Schwingenschlogl}},
  \bibinfo{journal}{2D Mater.} \textbf{\bibinfo{volume}{3}},
  \bibinfo{pages}{035012} (\bibinfo{year}{2016}).

\bibitem[{\citenamefont{Gurel and Salmankurt}(2021)}]{si-gu21}
\bibinfo{author}{\bibfnamefont{H.~H.} \bibnamefont{Gurel}} \bibnamefont{and}
  \bibinfo{author}{\bibfnamefont{B.}~\bibnamefont{Salmankurt}},
  \bibinfo{journal}{Biosensors} \textbf{\bibinfo{volume}{11}},
  \bibinfo{pages}{59} (\bibinfo{year}{2021}).

\bibitem[{\citenamefont{Guo et~al.}(2021)\citenamefont{Guo, Liu, Bai, Chen, and
  Qu}}]{si-gu21b}
\bibinfo{author}{\bibfnamefont{Q.}~\bibnamefont{Guo}},
  \bibinfo{author}{\bibfnamefont{J.}~\bibnamefont{Liu}},
  \bibinfo{author}{\bibfnamefont{C.}~\bibnamefont{Bai}},
  \bibinfo{author}{\bibfnamefont{N.}~\bibnamefont{Chen}}, \bibnamefont{and}
  \bibinfo{author}{\bibfnamefont{L.}~\bibnamefont{Qu}}, \bibinfo{journal}{ACS
  Nano} \textbf{\bibinfo{volume}{15}}, \bibinfo{pages}{16533}
  (\bibinfo{year}{2021}).

\bibitem[{\citenamefont{Do et~al.}(2019)\citenamefont{Do, Gumbs, Shih, Huang,
  Chiu, Chen, and Lin}}]{si-do19}
\bibinfo{author}{\bibfnamefont{T.-N.} \bibnamefont{Do}},
  \bibinfo{author}{\bibfnamefont{G.}~\bibnamefont{Gumbs}},
  \bibinfo{author}{\bibfnamefont{P.-H.} \bibnamefont{Shih}},
  \bibinfo{author}{\bibfnamefont{D.}~\bibnamefont{Huang}},
  \bibinfo{author}{\bibfnamefont{C.-W.} \bibnamefont{Chiu}},
  \bibinfo{author}{\bibfnamefont{C.-Y.} \bibnamefont{Chen}}, \bibnamefont{and}
  \bibinfo{author}{\bibfnamefont{M.-F.} \bibnamefont{Lin}},
  \bibinfo{journal}{Sci Rep.} \textbf{\bibinfo{volume}{9}},
  \bibinfo{pages}{624} (\bibinfo{year}{2019}).

\bibitem[{\citenamefont{Cahangirov et~al.}(2009)\citenamefont{Cahangirov,
  Topsakal, Akturk, Sahin, and Ciraci}}]{si-ca09}
\bibinfo{author}{\bibfnamefont{S.}~\bibnamefont{Cahangirov}},
  \bibinfo{author}{\bibfnamefont{M.}~\bibnamefont{Topsakal}},
  \bibinfo{author}{\bibfnamefont{E.}~\bibnamefont{Akturk}},
  \bibinfo{author}{\bibfnamefont{H.}~\bibnamefont{Sahin}}, \bibnamefont{and}
  \bibinfo{author}{\bibfnamefont{S.}~\bibnamefont{Ciraci}},
  \bibinfo{journal}{Phys. Rev. Lett.} \textbf{\bibinfo{volume}{102}},
  \bibinfo{pages}{236804} (\bibinfo{year}{2009}).

\bibitem[{\citenamefont{Trivedi et~al.}(2014)\citenamefont{Trivedi, Srivastava,
  and Kurchania}}]{si-tr14}
\bibinfo{author}{\bibfnamefont{S.}~\bibnamefont{Trivedi}},
  \bibinfo{author}{\bibfnamefont{A.}~\bibnamefont{Srivastava}},
  \bibnamefont{and}
  \bibinfo{author}{\bibfnamefont{R.}~\bibnamefont{Kurchania}},
  \bibinfo{journal}{J. Comp. Theor. Nanosci} \textbf{\bibinfo{volume}{11}},
  \bibinfo{pages}{781} (\bibinfo{year}{2014}).

\bibitem[{\citenamefont{Kumar and Suryanarayana}(2020)}]{si-ku20}
\bibinfo{author}{\bibfnamefont{S.}~\bibnamefont{Kumar}} \bibnamefont{and}
  \bibinfo{author}{\bibfnamefont{P.}~\bibnamefont{Suryanarayana}},
  \bibinfo{journal}{Nanotech.} \textbf{\bibinfo{volume}{31}},
  \bibinfo{pages}{43LT01} (\bibinfo{year}{2020}).

\bibitem[{\citenamefont{Banerjee and Suryanarayana}(2016)}]{si-ba16}
\bibinfo{author}{\bibfnamefont{A.~S.} \bibnamefont{Banerjee}} \bibnamefont{and}
  \bibinfo{author}{\bibfnamefont{P.}~\bibnamefont{Suryanarayana}},
  \bibinfo{journal}{J. Mech. Phys. Solids} \textbf{\bibinfo{volume}{96}},
  \bibinfo{pages}{605} (\bibinfo{year}{2016}).

\bibitem[{\citenamefont{Pizzochero et~al.}(2019)\citenamefont{Pizzochero,
  Bonfanti, and Martinazzo}}]{si-pi19}
\bibinfo{author}{\bibfnamefont{M.}~\bibnamefont{Pizzochero}},
  \bibinfo{author}{\bibfnamefont{M.}~\bibnamefont{Bonfanti}}, \bibnamefont{and}
  \bibinfo{author}{\bibfnamefont{R.}~\bibnamefont{Martinazzo}},
  \bibinfo{journal}{Phys. Chem. Chem. Phys.} \textbf{\bibinfo{volume}{21}},
  \bibinfo{pages}{26342} (\bibinfo{year}{2019}).

\bibitem[{\citenamefont{Zhao}(2012)}]{si-zh12}
\bibinfo{author}{\bibfnamefont{H.}~\bibnamefont{Zhao}}, \bibinfo{journal}{Phys.
  Lett. A} \textbf{\bibinfo{volume}{376}}, \bibinfo{pages}{3546}
  (\bibinfo{year}{2012}).

\bibitem[{\citenamefont{Mortazavi et~al.}(2017)\citenamefont{Mortazavi,
  Rahaman, Makaremi, Dianat, Cuniberti, and Rabczuk}}]{si-mo17}
\bibinfo{author}{\bibfnamefont{B.}~\bibnamefont{Mortazavi}},
  \bibinfo{author}{\bibfnamefont{O.}~\bibnamefont{Rahaman}},
  \bibinfo{author}{\bibfnamefont{M.}~\bibnamefont{Makaremi}},
  \bibinfo{author}{\bibfnamefont{A.}~\bibnamefont{Dianat}},
  \bibinfo{author}{\bibfnamefont{G.}~\bibnamefont{Cuniberti}},
  \bibnamefont{and} \bibinfo{author}{\bibfnamefont{T.}~\bibnamefont{Rabczuk}},
  \bibinfo{journal}{Physica E} \textbf{\bibinfo{volume}{87}},
  \bibinfo{pages}{228} (\bibinfo{year}{2017}).

\bibitem[{\citenamefont{Yoo et~al.}(2021)\citenamefont{Yoo, Lee, and
  Kang}}]{si-yo21}
\bibinfo{author}{\bibfnamefont{S.}~\bibnamefont{Yoo}},
  \bibinfo{author}{\bibfnamefont{B.}~\bibnamefont{Lee}}, \bibnamefont{and}
  \bibinfo{author}{\bibfnamefont{K.}~\bibnamefont{Kang}},
  \bibinfo{journal}{Nanotech.} \textbf{\bibinfo{volume}{32}},
  \bibinfo{pages}{295702} (\bibinfo{year}{2021}).

\bibitem[{\citenamefont{Zeng et~al.}(2018)\citenamefont{Zeng, Wu, Xu, Tao, Li,
  and Ouyang}}]{si-ze18}
\bibinfo{author}{\bibfnamefont{J.}~\bibnamefont{Zeng}},
  \bibinfo{author}{\bibfnamefont{M.}~\bibnamefont{Wu}},
  \bibinfo{author}{\bibfnamefont{B.}~\bibnamefont{Xu}},
  \bibinfo{author}{\bibfnamefont{S.}~\bibnamefont{Tao}},
  \bibinfo{author}{\bibfnamefont{X.}~\bibnamefont{Li}}, \bibnamefont{and}
  \bibinfo{author}{\bibfnamefont{C.}~\bibnamefont{Ouyang}},
  \bibinfo{journal}{J. Mater. Sci.} \textbf{\bibinfo{volume}{53}},
  \bibinfo{pages}{4306} (\bibinfo{year}{2018}).

\bibitem[{\citenamefont{Huang et~al.}(2015)\citenamefont{Huang, Gong, and
  Zeng}}]{si-hu15}
\bibinfo{author}{\bibfnamefont{L.-F.} \bibnamefont{Huang}},
  \bibinfo{author}{\bibfnamefont{P.-L.} \bibnamefont{Gong}}, \bibnamefont{and}
  \bibinfo{author}{\bibfnamefont{Z.}~\bibnamefont{Zeng}},
  \bibinfo{journal}{Phys. Rev. B} \textbf{\bibinfo{volume}{91}},
  \bibinfo{pages}{205433} (\bibinfo{year}{2015}).

\bibitem[{\citenamefont{Yan et~al.}(2013)\citenamefont{Yan, Stein, Schaefer,
  Wang, and Chou}}]{si-ya13}
\bibinfo{author}{\bibfnamefont{J.-A.} \bibnamefont{Yan}},
  \bibinfo{author}{\bibfnamefont{R.}~\bibnamefont{Stein}},
  \bibinfo{author}{\bibfnamefont{D.~M.} \bibnamefont{Schaefer}},
  \bibinfo{author}{\bibfnamefont{X.-Q.} \bibnamefont{Wang}}, \bibnamefont{and}
  \bibinfo{author}{\bibfnamefont{M.~Y.} \bibnamefont{Chou}},
  \bibinfo{journal}{Phys. Rev. B} \textbf{\bibinfo{volume}{88}},
  \bibinfo{pages}{121403} (\bibinfo{year}{2013}).

\bibitem[{\citenamefont{Peng et~al.}(2016{\natexlab{a}})\citenamefont{Peng,
  Zhang, Shao, Xu, Ni, Zhang, and Zhu}}]{si-pe16b}
\bibinfo{author}{\bibfnamefont{B.}~\bibnamefont{Peng}},
  \bibinfo{author}{\bibfnamefont{H.}~\bibnamefont{Zhang}},
  \bibinfo{author}{\bibfnamefont{H.}~\bibnamefont{Shao}},
  \bibinfo{author}{\bibfnamefont{Y.}~\bibnamefont{Xu}},
  \bibinfo{author}{\bibfnamefont{G.}~\bibnamefont{Ni}},
  \bibinfo{author}{\bibfnamefont{R.}~\bibnamefont{Zhang}}, \bibnamefont{and}
  \bibinfo{author}{\bibfnamefont{H.}~\bibnamefont{Zhu}},
  \bibinfo{journal}{Phys. Rev. B} \textbf{\bibinfo{volume}{94}},
  \bibinfo{pages}{245420} (\bibinfo{year}{2016}{\natexlab{a}}).

\bibitem[{\citenamefont{Peng et~al.}(2016{\natexlab{b}})\citenamefont{Peng,
  Zhang, Shao, Xu, Zhang, Lu, Zhang, and Zhu}}]{si-pe16}
\bibinfo{author}{\bibfnamefont{B.}~\bibnamefont{Peng}},
  \bibinfo{author}{\bibfnamefont{H.}~\bibnamefont{Zhang}},
  \bibinfo{author}{\bibfnamefont{H.}~\bibnamefont{Shao}},
  \bibinfo{author}{\bibfnamefont{Y.}~\bibnamefont{Xu}},
  \bibinfo{author}{\bibfnamefont{R.}~\bibnamefont{Zhang}},
  \bibinfo{author}{\bibfnamefont{H.}~\bibnamefont{Lu}},
  \bibinfo{author}{\bibfnamefont{D.~W.} \bibnamefont{Zhang}}, \bibnamefont{and}
  \bibinfo{author}{\bibfnamefont{H.}~\bibnamefont{Zhu}}, \bibinfo{journal}{ACS
  Appl. Mater. Interf.} \textbf{\bibinfo{volume}{8}}, \bibinfo{pages}{20977}
  (\bibinfo{year}{2016}{\natexlab{b}}).

\bibitem[{\citenamefont{Xie et~al.}(2014)\citenamefont{Xie, Hu, and
  Bao}}]{si-xi14}
\bibinfo{author}{\bibfnamefont{H.}~\bibnamefont{Xie}},
  \bibinfo{author}{\bibfnamefont{M.}~\bibnamefont{Hu}}, \bibnamefont{and}
  \bibinfo{author}{\bibfnamefont{H.}~\bibnamefont{Bao}},
  \bibinfo{journal}{Appl. Phys. Lett} \textbf{\bibinfo{volume}{104}},
  \bibinfo{pages}{131906} (\bibinfo{year}{2014}).

\bibitem[{\citenamefont{Gu and Yang}(2015)}]{si-gu15}
\bibinfo{author}{\bibfnamefont{X.}~\bibnamefont{Gu}} \bibnamefont{and}
  \bibinfo{author}{\bibfnamefont{R.}~\bibnamefont{Yang}}, \bibinfo{journal}{J.
  Appl. Phys.} \textbf{\bibinfo{volume}{117}}, \bibinfo{pages}{025102}
  (\bibinfo{year}{2015}).

\bibitem[{\citenamefont{Jaroch et~al.}(2021)\citenamefont{Jaroch, Krawiec, and
  Zdyb}}]{si-ja21}
\bibinfo{author}{\bibfnamefont{T.}~\bibnamefont{Jaroch}},
  \bibinfo{author}{\bibfnamefont{M.}~\bibnamefont{Krawiec}}, \bibnamefont{and}
  \bibinfo{author}{\bibfnamefont{R.}~\bibnamefont{Zdyb}}, \bibinfo{journal}{2D
  Mater.} \textbf{\bibinfo{volume}{8}}, \bibinfo{pages}{035038}
  (\bibinfo{year}{2021}).

\bibitem[{\citenamefont{Padilha and Pontes}(2015)}]{si-pa15}
\bibinfo{author}{\bibfnamefont{J.~E.} \bibnamefont{Padilha}} \bibnamefont{and}
  \bibinfo{author}{\bibfnamefont{R.~B.} \bibnamefont{Pontes}},
  \bibinfo{journal}{J. Phys. Chem. C} \textbf{\bibinfo{volume}{119}},
  \bibinfo{pages}{3818} (\bibinfo{year}{2015}).

\bibitem[{\citenamefont{Ipaves et~al.}(2022)\citenamefont{Ipaves, Justo, and
  Assali}}]{si-ip22}
\bibinfo{author}{\bibfnamefont{B.}~\bibnamefont{Ipaves}},
  \bibinfo{author}{\bibfnamefont{J.~F.} \bibnamefont{Justo}}, \bibnamefont{and}
  \bibinfo{author}{\bibfnamefont{L.~V.~C.} \bibnamefont{Assali}},
  \bibinfo{journal}{Phys. Chem. Chem. Phys.} \textbf{\bibinfo{volume}{24}},
  \bibinfo{pages}{8705} (\bibinfo{year}{2022}).

\bibitem[{\citenamefont{Yang et~al.}(2014)\citenamefont{Yang, Cahangirov,
  Cantarero, Rubio, and D'Agosta}}]{si-ya14b}
\bibinfo{author}{\bibfnamefont{K.}~\bibnamefont{Yang}},
  \bibinfo{author}{\bibfnamefont{S.}~\bibnamefont{Cahangirov}},
  \bibinfo{author}{\bibfnamefont{A.}~\bibnamefont{Cantarero}},
  \bibinfo{author}{\bibfnamefont{A.}~\bibnamefont{Rubio}}, \bibnamefont{and}
  \bibinfo{author}{\bibfnamefont{R.}~\bibnamefont{D'Agosta}},
  \bibinfo{journal}{Phys. Rev. B} \textbf{\bibinfo{volume}{89}},
  \bibinfo{pages}{125403} (\bibinfo{year}{2014}).

\bibitem[{\citenamefont{Berdiyorov and Peeters}(2014)}]{si-be14}
\bibinfo{author}{\bibfnamefont{G.~R.} \bibnamefont{Berdiyorov}}
  \bibnamefont{and} \bibinfo{author}{\bibfnamefont{F.~M.}
  \bibnamefont{Peeters}}, \bibinfo{journal}{RSC Adv.}
  \textbf{\bibinfo{volume}{4}}, \bibinfo{pages}{1133} (\bibinfo{year}{2014}).

\bibitem[{\citenamefont{Das and Sarkar}(2018)}]{si-da18}
\bibinfo{author}{\bibfnamefont{D.~K.} \bibnamefont{Das}} \bibnamefont{and}
  \bibinfo{author}{\bibfnamefont{J.}~\bibnamefont{Sarkar}},
  \bibinfo{journal}{J. Appl. Phys.} \textbf{\bibinfo{volume}{123}},
  \bibinfo{pages}{044304} (\bibinfo{year}{2018}).

\bibitem[{\citenamefont{Hu et~al.}(2013)\citenamefont{Hu, Zhang, and
  Poulikakos}}]{si-hu13b}
\bibinfo{author}{\bibfnamefont{M.}~\bibnamefont{Hu}},
  \bibinfo{author}{\bibfnamefont{X.}~\bibnamefont{Zhang}}, \bibnamefont{and}
  \bibinfo{author}{\bibfnamefont{D.}~\bibnamefont{Poulikakos}},
  \bibinfo{journal}{Phys. Rev. B} \textbf{\bibinfo{volume}{87}},
  \bibinfo{pages}{195417} (\bibinfo{year}{2013}).

\bibitem[{\citenamefont{Ince and Erkoc}(2011)}]{si-in11}
\bibinfo{author}{\bibfnamefont{A.}~\bibnamefont{Ince}} \bibnamefont{and}
  \bibinfo{author}{\bibfnamefont{S.}~\bibnamefont{Erkoc}},
  \bibinfo{journal}{Comp. Mater Sci.} \textbf{\bibinfo{volume}{50}},
  \bibinfo{pages}{865} (\bibinfo{year}{2011}).

\bibitem[{\citenamefont{Long et~al.}(2023)\citenamefont{Long, Tuan, Thuy, Ho,
  and Huy}}]{si-lo23}
\bibinfo{author}{\bibfnamefont{N.~T.} \bibnamefont{Long}},
  \bibinfo{author}{\bibfnamefont{T.~Q.} \bibnamefont{Tuan}},
  \bibinfo{author}{\bibfnamefont{D.~N.~A.} \bibnamefont{Thuy}},
  \bibinfo{author}{\bibfnamefont{Q.~D.} \bibnamefont{Ho}}, \bibnamefont{and}
  \bibinfo{author}{\bibfnamefont{H.~A.} \bibnamefont{Huy}},
  \bibinfo{journal}{Molec. Simul.} \textbf{\bibinfo{volume}{49}},
  \bibinfo{pages}{655} (\bibinfo{year}{2023}).

\bibitem[{\citenamefont{Min et~al.}(2018)\citenamefont{Min, Yoon, and
  Lim}}]{si-mi18}
\bibinfo{author}{\bibfnamefont{T.~K.} \bibnamefont{Min}},
  \bibinfo{author}{\bibfnamefont{T.~L.} \bibnamefont{Yoon}}, \bibnamefont{and}
  \bibinfo{author}{\bibfnamefont{T.~L.} \bibnamefont{Lim}},
  \bibinfo{journal}{Mater. Res. Express} \textbf{\bibinfo{volume}{5}},
  \bibinfo{pages}{065054} (\bibinfo{year}{2018}).

\bibitem[{\citenamefont{Pei et~al.}(2014)\citenamefont{Pei, Sha, Zhang, and
  Zhang}}]{si-pe14}
\bibinfo{author}{\bibfnamefont{Q.-X.} \bibnamefont{Pei}},
  \bibinfo{author}{\bibfnamefont{Z.-D.} \bibnamefont{Sha}},
  \bibinfo{author}{\bibfnamefont{Y.-Y.} \bibnamefont{Zhang}}, \bibnamefont{and}
  \bibinfo{author}{\bibfnamefont{Y.-W.} \bibnamefont{Zhang}},
  \bibinfo{journal}{J. Appl. Phys.} \textbf{\bibinfo{volume}{115}},
  \bibinfo{pages}{023519} (\bibinfo{year}{2014}).

\bibitem[{\citenamefont{Rouhi et~al.}(2019)\citenamefont{Rouhi, Pourmirzaagha,
  and Farzin}}]{si-ro19}
\bibinfo{author}{\bibfnamefont{S.}~\bibnamefont{Rouhi}},
  \bibinfo{author}{\bibfnamefont{H.}~\bibnamefont{Pourmirzaagha}},
  \bibnamefont{and} \bibinfo{author}{\bibfnamefont{A.}~\bibnamefont{Farzin}},
  \bibinfo{journal}{Mater. Res. Express} \textbf{\bibinfo{volume}{6}},
  \bibinfo{pages}{085004} (\bibinfo{year}{2019}).

\bibitem[{\citenamefont{Wang et~al.}(2015)\citenamefont{Wang, Feng, and
  Ruan}}]{si-wa15}
\bibinfo{author}{\bibfnamefont{Z.}~\bibnamefont{Wang}},
  \bibinfo{author}{\bibfnamefont{T.}~\bibnamefont{Feng}}, \bibnamefont{and}
  \bibinfo{author}{\bibfnamefont{X.}~\bibnamefont{Ruan}}, \bibinfo{journal}{J.
  Appl. Phys.} \textbf{\bibinfo{volume}{117}}, \bibinfo{pages}{084317}
  (\bibinfo{year}{2015}).

\bibitem[{\citenamefont{Rouhi}(2017)}]{si-ro17}
\bibinfo{author}{\bibfnamefont{S.}~\bibnamefont{Rouhi}},
  \bibinfo{journal}{Comp. Mater, Sci.} \textbf{\bibinfo{volume}{131}},
  \bibinfo{pages}{275} (\bibinfo{year}{2017}).

\bibitem[{\citenamefont{Feynman}(1972)}]{fe72}
\bibinfo{author}{\bibfnamefont{R.~P.} \bibnamefont{Feynman}},
  \emph{\bibinfo{title}{Statistical Mechanics}}
  (\bibinfo{publisher}{Addison-Wesley}, \bibinfo{address}{New York},
  \bibinfo{year}{1972}).

\bibitem[{\citenamefont{Gillan}(1988)}]{gi88}
\bibinfo{author}{\bibfnamefont{M.~J.} \bibnamefont{Gillan}},
  \bibinfo{journal}{Phil. Mag. A} \textbf{\bibinfo{volume}{58}},
  \bibinfo{pages}{257} (\bibinfo{year}{1988}).

\bibitem[{\citenamefont{Herrero and Ram\'irez}(2014)}]{he14}
\bibinfo{author}{\bibfnamefont{C.~P.} \bibnamefont{Herrero}} \bibnamefont{and}
  \bibinfo{author}{\bibfnamefont{R.}~\bibnamefont{Ram\'irez}},
  \bibinfo{journal}{J. Phys.: Condens. Matter} \textbf{\bibinfo{volume}{26}},
  \bibinfo{pages}{233201} (\bibinfo{year}{2014}).

\bibitem[{\citenamefont{Ceperley}(1995)}]{ce95}
\bibinfo{author}{\bibfnamefont{D.~M.} \bibnamefont{Ceperley}},
  \bibinfo{journal}{Rev. Mod. Phys.} \textbf{\bibinfo{volume}{67}},
  \bibinfo{pages}{279} (\bibinfo{year}{1995}).

\bibitem[{\citenamefont{Herrero and Ram\'irez}(2016)}]{he16}
\bibinfo{author}{\bibfnamefont{C.~P.} \bibnamefont{Herrero}} \bibnamefont{and}
  \bibinfo{author}{\bibfnamefont{R.}~\bibnamefont{Ram\'irez}},
  \bibinfo{journal}{J. Chem. Phys.} \textbf{\bibinfo{volume}{145}},
  \bibinfo{pages}{224701} (\bibinfo{year}{2016}).

\bibitem[{\citenamefont{Brito et~al.}(2015)\citenamefont{Brito, Candido, Hai,
  and Peeters}}]{br15}
\bibinfo{author}{\bibfnamefont{B.~G.~A.} \bibnamefont{Brito}},
  \bibinfo{author}{\bibfnamefont{L.}~\bibnamefont{Candido}},
  \bibinfo{author}{\bibfnamefont{G.~Q.} \bibnamefont{Hai}}, \bibnamefont{and}
  \bibinfo{author}{\bibfnamefont{F.~M.} \bibnamefont{Peeters}},
  \bibinfo{journal}{Phys. Rev. B} \textbf{\bibinfo{volume}{92}},
  \bibinfo{pages}{195416} (\bibinfo{year}{2015}).

\bibitem[{\citenamefont{Herrero and Ram\'irez}(2022)}]{he22}
\bibinfo{author}{\bibfnamefont{C.~P.} \bibnamefont{Herrero}} \bibnamefont{and}
  \bibinfo{author}{\bibfnamefont{R.}~\bibnamefont{Ram\'irez}},
  \bibinfo{journal}{J. Phys. Chem. Solids} \textbf{\bibinfo{volume}{171}},
  \bibinfo{pages}{110980} (\bibinfo{year}{2022}).

\bibitem[{\citenamefont{Brito et~al.}(2022)\citenamefont{Brito, Candido,
  Teixeira~Rabelo, and Hai}}]{br22}
\bibinfo{author}{\bibfnamefont{B.~G.~A.} \bibnamefont{Brito}},
  \bibinfo{author}{\bibfnamefont{L.}~\bibnamefont{Candido}},
  \bibinfo{author}{\bibfnamefont{J.~N.} \bibnamefont{Teixeira~Rabelo}},
  \bibnamefont{and} \bibinfo{author}{\bibfnamefont{G.~Q.} \bibnamefont{Hai}},
  \bibinfo{journal}{Comp. Condens. Matter} \textbf{\bibinfo{volume}{31}},
  \bibinfo{pages}{e00660} (\bibinfo{year}{2022}).

\bibitem[{\citenamefont{Herrero and Ram\'irez}(2018)}]{he18b}
\bibinfo{author}{\bibfnamefont{C.~P.} \bibnamefont{Herrero}} \bibnamefont{and}
  \bibinfo{author}{\bibfnamefont{R.}~\bibnamefont{Ram\'irez}},
  \bibinfo{journal}{Phys. Rev. B} \textbf{\bibinfo{volume}{97}},
  \bibinfo{pages}{195433} (\bibinfo{year}{2018}).

\bibitem[{\citenamefont{Herrero and Ram\'irez}(2020{\natexlab{a}})}]{he20c}
\bibinfo{author}{\bibfnamefont{C.~P.} \bibnamefont{Herrero}} \bibnamefont{and}
  \bibinfo{author}{\bibfnamefont{R.}~\bibnamefont{Ram\'irez}},
  \bibinfo{journal}{Eur. Phys. J. B} \textbf{\bibinfo{volume}{93}},
  \bibinfo{pages}{146} (\bibinfo{year}{2020}{\natexlab{a}}).

\bibitem[{\citenamefont{Noya et~al.}(1996)\citenamefont{Noya, Herrero, and
  Ram\'{\i}rez}}]{no96}
\bibinfo{author}{\bibfnamefont{J.~C.} \bibnamefont{Noya}},
  \bibinfo{author}{\bibfnamefont{C.~P.} \bibnamefont{Herrero}},
  \bibnamefont{and}
  \bibinfo{author}{\bibfnamefont{R.}~\bibnamefont{Ram\'{\i}rez}},
  \bibinfo{journal}{Phys. Rev. B} \textbf{\bibinfo{volume}{53}},
  \bibinfo{pages}{9869} (\bibinfo{year}{1996}).

\bibitem[{\citenamefont{Cazorla and Boronat}(2017)}]{ca17}
\bibinfo{author}{\bibfnamefont{C.}~\bibnamefont{Cazorla}} \bibnamefont{and}
  \bibinfo{author}{\bibfnamefont{J.}~\bibnamefont{Boronat}},
  \bibinfo{journal}{Rev. Mod. Phys.} \textbf{\bibinfo{volume}{89}},
  \bibinfo{pages}{035003} (\bibinfo{year}{2017}).

\bibitem[{\citenamefont{Porezag et~al.}(1995)\citenamefont{Porezag, Frauenheim,
  K\"ohler, Seifert, and Kaschner}}]{po95}
\bibinfo{author}{\bibfnamefont{D.}~\bibnamefont{Porezag}},
  \bibinfo{author}{\bibfnamefont{T.}~\bibnamefont{Frauenheim}},
  \bibinfo{author}{\bibfnamefont{T.}~\bibnamefont{K\"ohler}},
  \bibinfo{author}{\bibfnamefont{G.}~\bibnamefont{Seifert}}, \bibnamefont{and}
  \bibinfo{author}{\bibfnamefont{R.}~\bibnamefont{Kaschner}},
  \bibinfo{journal}{Phys. Rev. B} \textbf{\bibinfo{volume}{51}},
  \bibinfo{pages}{12947} (\bibinfo{year}{1995}).

\bibitem[{\citenamefont{Gutierrez et~al.}(1996)\citenamefont{Gutierrez,
  Frauenheim, K\"ohler, and Seifert}}]{gu96}
\bibinfo{author}{\bibfnamefont{R.}~\bibnamefont{Gutierrez}},
  \bibinfo{author}{\bibfnamefont{T.}~\bibnamefont{Frauenheim}},
  \bibinfo{author}{\bibfnamefont{T.}~\bibnamefont{K\"ohler}}, \bibnamefont{and}
  \bibinfo{author}{\bibfnamefont{G.}~\bibnamefont{Seifert}},
  \bibinfo{journal}{J. Mater. Chem.} \textbf{\bibinfo{volume}{6}},
  \bibinfo{pages}{1657} (\bibinfo{year}{1996}).

\bibitem[{\citenamefont{Goringe et~al.}(1997)\citenamefont{Goringe, Bowler, and
  Hern\'andez}}]{go97}
\bibinfo{author}{\bibfnamefont{C.~M.} \bibnamefont{Goringe}},
  \bibinfo{author}{\bibfnamefont{D.~R.} \bibnamefont{Bowler}},
  \bibnamefont{and}
  \bibinfo{author}{\bibfnamefont{E.}~\bibnamefont{Hern\'andez}},
  \bibinfo{journal}{Rep. Prog. Phys.} \textbf{\bibinfo{volume}{60}},
  \bibinfo{pages}{1447} (\bibinfo{year}{1997}).

\bibitem[{\citenamefont{Frauenheim et~al.}(1995)\citenamefont{Frauenheim,
  Weich, Kohler, Uhlmann, Porezag, and Seifert}}]{fr95}
\bibinfo{author}{\bibfnamefont{T.}~\bibnamefont{Frauenheim}},
  \bibinfo{author}{\bibfnamefont{F.}~\bibnamefont{Weich}},
  \bibinfo{author}{\bibfnamefont{T.}~\bibnamefont{Kohler}},
  \bibinfo{author}{\bibfnamefont{S.}~\bibnamefont{Uhlmann}},
  \bibinfo{author}{\bibfnamefont{D.}~\bibnamefont{Porezag}}, \bibnamefont{and}
  \bibinfo{author}{\bibfnamefont{G.}~\bibnamefont{Seifert}},
  \bibinfo{journal}{Phys. Rev. B} \textbf{\bibinfo{volume}{52}},
  \bibinfo{pages}{11492} (\bibinfo{year}{1995}).

\bibitem[{\citenamefont{Klein et~al.}(1999{\natexlab{a}})\citenamefont{Klein,
  Urbassek, and Frauenheim}}]{si-kl99}
\bibinfo{author}{\bibfnamefont{P.}~\bibnamefont{Klein}},
  \bibinfo{author}{\bibfnamefont{H.~M.} \bibnamefont{Urbassek}},
  \bibnamefont{and}
  \bibinfo{author}{\bibfnamefont{T.}~\bibnamefont{Frauenheim}},
  \bibinfo{journal}{Comp. Mater. Sci.} \textbf{\bibinfo{volume}{13}},
  \bibinfo{pages}{252} (\bibinfo{year}{1999}{\natexlab{a}}).

\bibitem[{\citenamefont{Klein et~al.}(1999{\natexlab{b}})\citenamefont{Klein,
  Urbassek, and Frauenheim}}]{si-kl99b}
\bibinfo{author}{\bibfnamefont{P.}~\bibnamefont{Klein}},
  \bibinfo{author}{\bibfnamefont{H.~M.} \bibnamefont{Urbassek}},
  \bibnamefont{and}
  \bibinfo{author}{\bibfnamefont{T.}~\bibnamefont{Frauenheim}},
  \bibinfo{journal}{Phys. Rev. B} \textbf{\bibinfo{volume}{60}},
  \bibinfo{pages}{5478} (\bibinfo{year}{1999}{\natexlab{b}}).

\bibitem[{\citenamefont{Kaczmarski et~al.}(2005)\citenamefont{Kaczmarski,
  Bedoya-Martinez, and Hernandez}}]{sc-ka05}
\bibinfo{author}{\bibfnamefont{M.}~\bibnamefont{Kaczmarski}},
  \bibinfo{author}{\bibfnamefont{O.~N.} \bibnamefont{Bedoya-Martinez}},
  \bibnamefont{and} \bibinfo{author}{\bibfnamefont{E.~R.}
  \bibnamefont{Hernandez}}, \bibinfo{journal}{Phys. Rev. Lett.}
  \textbf{\bibinfo{volume}{94}}, \bibinfo{pages}{095701}
  (\bibinfo{year}{2005}).

\bibitem[{\citenamefont{Shevlin et~al.}(2001)\citenamefont{Shevlin, Fisher, and
  Hernandez}}]{sc-sh01}
\bibinfo{author}{\bibfnamefont{S.~A.} \bibnamefont{Shevlin}},
  \bibinfo{author}{\bibfnamefont{A.~J.} \bibnamefont{Fisher}},
  \bibnamefont{and}
  \bibinfo{author}{\bibfnamefont{E.}~\bibnamefont{Hernandez}},
  \bibinfo{journal}{Phys. Rev. B} \textbf{\bibinfo{volume}{63}},
  \bibinfo{pages}{195306} (\bibinfo{year}{2001}).

\bibitem[{\citenamefont{Ram\'irez et~al.}(2008)\citenamefont{Ram\'irez,
  Herrero, Hern\'andez, and Cardona}}]{ra08}
\bibinfo{author}{\bibfnamefont{R.}~\bibnamefont{Ram\'irez}},
  \bibinfo{author}{\bibfnamefont{C.~P.} \bibnamefont{Herrero}},
  \bibinfo{author}{\bibfnamefont{E.~R.} \bibnamefont{Hern\'andez}},
  \bibnamefont{and} \bibinfo{author}{\bibfnamefont{M.}~\bibnamefont{Cardona}},
  \bibinfo{journal}{Phys. Rev. B} \textbf{\bibinfo{volume}{77}},
  \bibinfo{pages}{045210} (\bibinfo{year}{2008}).

\bibitem[{\citenamefont{Marton\'ak et~al.}(2002)\citenamefont{Marton\'ak,
  Colombo, Molteni, and Parrinello}}]{si-ma02}
\bibinfo{author}{\bibfnamefont{R.}~\bibnamefont{Marton\'ak}},
  \bibinfo{author}{\bibfnamefont{L.}~\bibnamefont{Colombo}},
  \bibinfo{author}{\bibfnamefont{C.}~\bibnamefont{Molteni}}, \bibnamefont{and}
  \bibinfo{author}{\bibfnamefont{M.}~\bibnamefont{Parrinello}},
  \bibinfo{journal}{J. Chem. Phys.} \textbf{\bibinfo{volume}{117}},
  \bibinfo{pages}{11329} (\bibinfo{year}{2002}).

\bibitem[{\citenamefont{Alippi et~al.}(2001)\citenamefont{Alippi, Colombo,
  Ruggerone, Sieck, Seifert, and Frauenheim}}]{si-al01}
\bibinfo{author}{\bibfnamefont{P.}~\bibnamefont{Alippi}},
  \bibinfo{author}{\bibfnamefont{L.}~\bibnamefont{Colombo}},
  \bibinfo{author}{\bibfnamefont{P.}~\bibnamefont{Ruggerone}},
  \bibinfo{author}{\bibfnamefont{A.}~\bibnamefont{Sieck}},
  \bibinfo{author}{\bibfnamefont{G.}~\bibnamefont{Seifert}}, \bibnamefont{and}
  \bibinfo{author}{\bibfnamefont{T.}~\bibnamefont{Frauenheim}},
  \bibinfo{journal}{Phys. Rev. B} \textbf{\bibinfo{volume}{64}},
  \bibinfo{pages}{075207} (\bibinfo{year}{2001}).

\bibitem[{\citenamefont{Cogoni et~al.}(2005)\citenamefont{Cogoni, Uberuaga,
  Voter, and Colombo}}]{si-co05}
\bibinfo{author}{\bibfnamefont{M.}~\bibnamefont{Cogoni}},
  \bibinfo{author}{\bibfnamefont{B.}~\bibnamefont{Uberuaga}},
  \bibinfo{author}{\bibfnamefont{A.}~\bibnamefont{Voter}}, \bibnamefont{and}
  \bibinfo{author}{\bibfnamefont{L.}~\bibnamefont{Colombo}},
  \bibinfo{journal}{Phys. Rev. B} \textbf{\bibinfo{volume}{71}},
  \bibinfo{pages}{121203} (\bibinfo{year}{2005}).

\bibitem[{\citenamefont{Ram\'irez and Herrero}(2017)}]{ra17}
\bibinfo{author}{\bibfnamefont{R.}~\bibnamefont{Ram\'irez}} \bibnamefont{and}
  \bibinfo{author}{\bibfnamefont{C.~P.} \bibnamefont{Herrero}},
  \bibinfo{journal}{Phys. Rev. B} \textbf{\bibinfo{volume}{95}},
  \bibinfo{pages}{045423} (\bibinfo{year}{2017}).

\bibitem[{\citenamefont{Fournier and Barbetta}(2008)}]{fo08}
\bibinfo{author}{\bibfnamefont{J.-B.} \bibnamefont{Fournier}} \bibnamefont{and}
  \bibinfo{author}{\bibfnamefont{C.}~\bibnamefont{Barbetta}},
  \bibinfo{journal}{Phys. Rev. Lett.} \textbf{\bibinfo{volume}{100}},
  \bibinfo{pages}{078103} (\bibinfo{year}{2008}).

\bibitem[{\citenamefont{Shiba et~al.}(2016)\citenamefont{Shiba, Noguchi, and
  Fournier}}]{sh16}
\bibinfo{author}{\bibfnamefont{H.}~\bibnamefont{Shiba}},
  \bibinfo{author}{\bibfnamefont{H.}~\bibnamefont{Noguchi}}, \bibnamefont{and}
  \bibinfo{author}{\bibfnamefont{J.-B.} \bibnamefont{Fournier}},
  \bibinfo{journal}{Soft Matter} \textbf{\bibinfo{volume}{12}},
  \bibinfo{pages}{2373} (\bibinfo{year}{2016}).

\bibitem[{\citenamefont{Martyna et~al.}(1996)\citenamefont{Martyna, Tuckerman,
  Tobias, and Klein}}]{ma96}
\bibinfo{author}{\bibfnamefont{G.~J.} \bibnamefont{Martyna}},
  \bibinfo{author}{\bibfnamefont{M.~E.} \bibnamefont{Tuckerman}},
  \bibinfo{author}{\bibfnamefont{D.~J.} \bibnamefont{Tobias}},
  \bibnamefont{and} \bibinfo{author}{\bibfnamefont{M.~L.} \bibnamefont{Klein}},
  \bibinfo{journal}{Mol. Phys.} \textbf{\bibinfo{volume}{87}},
  \bibinfo{pages}{1117} (\bibinfo{year}{1996}).

\bibitem[{\citenamefont{Ram\'irez and Herrero}(2020)}]{ra20}
\bibinfo{author}{\bibfnamefont{R.}~\bibnamefont{Ram\'irez}} \bibnamefont{and}
  \bibinfo{author}{\bibfnamefont{C.~P.} \bibnamefont{Herrero}},
  \bibinfo{journal}{Phys. Rev. B} \textbf{\bibinfo{volume}{101}},
  \bibinfo{pages}{235436} (\bibinfo{year}{2020}).

\bibitem[{\citenamefont{Tuckerman}(2010)}]{tu10}
\bibinfo{author}{\bibfnamefont{M.~E.} \bibnamefont{Tuckerman}},
  \emph{\bibinfo{title}{Statistical Mechanics: Theory and Molecular
  Simulation}} (\bibinfo{publisher}{Oxford University Press},
  \bibinfo{address}{Oxford}, \bibinfo{year}{2010}).

\bibitem[{\citenamefont{Herman et~al.}(1982)\citenamefont{Herman, Bruskin, and
  Berne}}]{he82}
\bibinfo{author}{\bibfnamefont{M.~F.} \bibnamefont{Herman}},
  \bibinfo{author}{\bibfnamefont{E.~J.} \bibnamefont{Bruskin}},
  \bibnamefont{and} \bibinfo{author}{\bibfnamefont{B.~J.} \bibnamefont{Berne}},
  \bibinfo{journal}{J. Chem. Phys.} \textbf{\bibinfo{volume}{76}},
  \bibinfo{pages}{5150} (\bibinfo{year}{1982}).

\bibitem[{\citenamefont{Herrero et~al.}(2024)\citenamefont{Herrero, Ram\'irez,
  and Herrero-Saboya}}]{sc-he24}
\bibinfo{author}{\bibfnamefont{C.~P.} \bibnamefont{Herrero}},
  \bibinfo{author}{\bibfnamefont{R.}~\bibnamefont{Ram\'irez}},
  \bibnamefont{and}
  \bibinfo{author}{\bibfnamefont{G.}~\bibnamefont{Herrero-Saboya}},
  \bibinfo{journal}{Phys. Rev. B} \textbf{\bibinfo{volume}{109}},
  \bibinfo{pages}{104112} (\bibinfo{year}{2024}).

\bibitem[{\citenamefont{Sahin et~al.}(2009)\citenamefont{Sahin, Cahangirov,
  Topsakal, Bekaroglu, Akturk, Senger, and Ciraci}}]{si-sa09}
\bibinfo{author}{\bibfnamefont{H.}~\bibnamefont{Sahin}},
  \bibinfo{author}{\bibfnamefont{S.}~\bibnamefont{Cahangirov}},
  \bibinfo{author}{\bibfnamefont{M.}~\bibnamefont{Topsakal}},
  \bibinfo{author}{\bibfnamefont{E.}~\bibnamefont{Bekaroglu}},
  \bibinfo{author}{\bibfnamefont{E.}~\bibnamefont{Akturk}},
  \bibinfo{author}{\bibfnamefont{R.~T.} \bibnamefont{Senger}},
  \bibnamefont{and} \bibinfo{author}{\bibfnamefont{S.}~\bibnamefont{Ciraci}},
  \bibinfo{journal}{Phys. Rev. B} \textbf{\bibinfo{volume}{80}},
  \bibinfo{pages}{155453} (\bibinfo{year}{2009}).

\bibitem[{\citenamefont{Scalise et~al.}(2013)\citenamefont{Scalise, Houssa,
  Pourtois, van~den Broek, Afanas'ev, and Stesmans}}]{si-sc13}
\bibinfo{author}{\bibfnamefont{E.}~\bibnamefont{Scalise}},
  \bibinfo{author}{\bibfnamefont{M.}~\bibnamefont{Houssa}},
  \bibinfo{author}{\bibfnamefont{G.}~\bibnamefont{Pourtois}},
  \bibinfo{author}{\bibfnamefont{B.}~\bibnamefont{van~den Broek}},
  \bibinfo{author}{\bibfnamefont{V.}~\bibnamefont{Afanas'ev}},
  \bibnamefont{and} \bibinfo{author}{\bibfnamefont{A.}~\bibnamefont{Stesmans}},
  \bibinfo{journal}{Nano Res.} \textbf{\bibinfo{volume}{6}},
  \bibinfo{pages}{19} (\bibinfo{year}{2013}).

\bibitem[{\citenamefont{Zhang et~al.}(2014)\citenamefont{Zhang, Xie, Hu, Bao,
  Yue, Qin, and Su}}]{si-zh14}
\bibinfo{author}{\bibfnamefont{X.}~\bibnamefont{Zhang}},
  \bibinfo{author}{\bibfnamefont{H.}~\bibnamefont{Xie}},
  \bibinfo{author}{\bibfnamefont{M.}~\bibnamefont{Hu}},
  \bibinfo{author}{\bibfnamefont{H.}~\bibnamefont{Bao}},
  \bibinfo{author}{\bibfnamefont{S.}~\bibnamefont{Yue}},
  \bibinfo{author}{\bibfnamefont{G.}~\bibnamefont{Qin}}, \bibnamefont{and}
  \bibinfo{author}{\bibfnamefont{G.}~\bibnamefont{Su}}, \bibinfo{journal}{Phys.
  Rev. B} \textbf{\bibinfo{volume}{89}}, \bibinfo{pages}{054310}
  (\bibinfo{year}{2014}).

\bibitem[{\citenamefont{Ram\'irez and B\"ohm}(1986)}]{ra86}
\bibinfo{author}{\bibfnamefont{R.}~\bibnamefont{Ram\'irez}} \bibnamefont{and}
  \bibinfo{author}{\bibfnamefont{M.~C.} \bibnamefont{B\"ohm}},
  \bibinfo{journal}{Inter. J. Quantum Chem.} \textbf{\bibinfo{volume}{30}},
  \bibinfo{pages}{391} (\bibinfo{year}{1986}).

\bibitem[{\citenamefont{Togo et~al.}(2023)\citenamefont{Togo, Chaput, Tadano,
  and Tanaka}}]{si-to23}
\bibinfo{author}{\bibfnamefont{A.}~\bibnamefont{Togo}},
  \bibinfo{author}{\bibfnamefont{L.}~\bibnamefont{Chaput}},
  \bibinfo{author}{\bibfnamefont{T.}~\bibnamefont{Tadano}}, \bibnamefont{and}
  \bibinfo{author}{\bibfnamefont{I.}~\bibnamefont{Tanaka}},
  \bibinfo{journal}{J. Phys.: Condens. Matter} \textbf{\bibinfo{volume}{35}},
  \bibinfo{pages}{353001} (\bibinfo{year}{2023}).

\bibitem[{\citenamefont{Wirtz and Rubio}(2004)}]{wi04}
\bibinfo{author}{\bibfnamefont{L.}~\bibnamefont{Wirtz}} \bibnamefont{and}
  \bibinfo{author}{\bibfnamefont{A.}~\bibnamefont{Rubio}},
  \bibinfo{journal}{Solid State Commun.} \textbf{\bibinfo{volume}{131}},
  \bibinfo{pages}{141} (\bibinfo{year}{2004}).

\bibitem[{\citenamefont{Mounet and Marzari}(2005)}]{mo05}
\bibinfo{author}{\bibfnamefont{N.}~\bibnamefont{Mounet}} \bibnamefont{and}
  \bibinfo{author}{\bibfnamefont{N.}~\bibnamefont{Marzari}},
  \bibinfo{journal}{Phys. Rev. B} \textbf{\bibinfo{volume}{71}},
  \bibinfo{pages}{205214} (\bibinfo{year}{2005}).

\bibitem[{\citenamefont{Yan et~al.}({2008})\citenamefont{Yan, Ruan, and
  Chou}}]{ya08}
\bibinfo{author}{\bibfnamefont{J.-A.} \bibnamefont{Yan}},
  \bibinfo{author}{\bibfnamefont{W.~Y.} \bibnamefont{Ruan}}, \bibnamefont{and}
  \bibinfo{author}{\bibfnamefont{M.~Y.} \bibnamefont{Chou}},
  \bibinfo{journal}{Phys. Rev. B} \textbf{\bibinfo{volume}{{77}}},
  \bibinfo{pages}{{125401}} (\bibinfo{year}{{2008}}).

\bibitem[{\citenamefont{Ram\'irez and Herrero}(2019)}]{ra19}
\bibinfo{author}{\bibfnamefont{R.}~\bibnamefont{Ram\'irez}} \bibnamefont{and}
  \bibinfo{author}{\bibfnamefont{C.~P.} \bibnamefont{Herrero}},
  \bibinfo{journal}{J. Chem. Phys.} \textbf{\bibinfo{volume}{151}},
  \bibinfo{pages}{224107} (\bibinfo{year}{2019}).

\bibitem[{\citenamefont{Ram\'{\i}rez et~al.}(2006)\citenamefont{Ram\'{\i}rez,
  Herrero, and Hern\'andez}}]{ra06}
\bibinfo{author}{\bibfnamefont{R.}~\bibnamefont{Ram\'{\i}rez}},
  \bibinfo{author}{\bibfnamefont{C.~P.} \bibnamefont{Herrero}},
  \bibnamefont{and} \bibinfo{author}{\bibfnamefont{E.~R.}
  \bibnamefont{Hern\'andez}}, \bibinfo{journal}{Phys. Rev. B}
  \textbf{\bibinfo{volume}{73}}, \bibinfo{pages}{245202}
  (\bibinfo{year}{2006}).

\bibitem[{\citenamefont{Garcia et~al.}(2011)\citenamefont{Garcia, de~Lima,
  Assali, and Justo}}]{si-ga11}
\bibinfo{author}{\bibfnamefont{J.~C.} \bibnamefont{Garcia}},
  \bibinfo{author}{\bibfnamefont{D.~B.} \bibnamefont{de~Lima}},
  \bibinfo{author}{\bibfnamefont{L.~V.~C.} \bibnamefont{Assali}},
  \bibnamefont{and} \bibinfo{author}{\bibfnamefont{J.~F.} \bibnamefont{Justo}},
  \bibinfo{journal}{J. Phys. Chem. C} \textbf{\bibinfo{volume}{115}},
  \bibinfo{pages}{13242} (\bibinfo{year}{2011}).

\bibitem[{\citenamefont{Kaltsas et~al.}(2014)\citenamefont{Kaltsas, Tsetseris,
  and Dimoulas}}]{si-ka14}
\bibinfo{author}{\bibfnamefont{D.}~\bibnamefont{Kaltsas}},
  \bibinfo{author}{\bibfnamefont{L.}~\bibnamefont{Tsetseris}},
  \bibnamefont{and} \bibinfo{author}{\bibfnamefont{A.}~\bibnamefont{Dimoulas}},
  \bibinfo{journal}{Appl. Surf. Sci.} \textbf{\bibinfo{volume}{291}},
  \bibinfo{pages}{93} (\bibinfo{year}{2014}).

\bibitem[{\citenamefont{Kittel}(2005)}]{ki05}
\bibinfo{author}{\bibfnamefont{C.}~\bibnamefont{Kittel}},
  \emph{\bibinfo{title}{Introduction to Solid State Physics}}
  (\bibinfo{publisher}{Wiley}, \bibinfo{address}{New York},
  \bibinfo{year}{2005}), \bibinfo{edition}{8th} ed.

\bibitem[{\citenamefont{Herrero and Ram\'irez}(2020{\natexlab{b}})}]{he20b}
\bibinfo{author}{\bibfnamefont{C.~P.} \bibnamefont{Herrero}} \bibnamefont{and}
  \bibinfo{author}{\bibfnamefont{R.}~\bibnamefont{Ram\'irez}},
  \bibinfo{journal}{Chem. Phys.} \textbf{\bibinfo{volume}{533}},
  \bibinfo{pages}{110737} (\bibinfo{year}{2020}{\natexlab{b}}).

\bibitem[{\citenamefont{Herrero and Ram\'irez}(2023)}]{he23}
\bibinfo{author}{\bibfnamefont{C.~P.} \bibnamefont{Herrero}} \bibnamefont{and}
  \bibinfo{author}{\bibfnamefont{R.}~\bibnamefont{Ram\'irez}},
  \bibinfo{journal}{Eur. Phys. J. B} \textbf{\bibinfo{volume}{96}},
  \bibinfo{pages}{147} (\bibinfo{year}{2023}).

\bibitem[{\citenamefont{Gao and Huang}(2014)}]{ga14}
\bibinfo{author}{\bibfnamefont{W.}~\bibnamefont{Gao}} \bibnamefont{and}
  \bibinfo{author}{\bibfnamefont{R.}~\bibnamefont{Huang}}, \bibinfo{journal}{J.
  Mech. Phys. Solids} \textbf{\bibinfo{volume}{66}}, \bibinfo{pages}{42}
  (\bibinfo{year}{2014}).

\bibitem[{\citenamefont{Masson and Prevot}(2023)}]{si-ma23}
\bibinfo{author}{\bibfnamefont{L.}~\bibnamefont{Masson}} \bibnamefont{and}
  \bibinfo{author}{\bibfnamefont{G.}~\bibnamefont{Prevot}},
  \bibinfo{journal}{Nanoscale Adv.} \textbf{\bibinfo{volume}{5}},
  \bibinfo{pages}{1574} (\bibinfo{year}{2023}).

\bibitem[{\citenamefont{Behroozi}(1996)}]{be96b}
\bibinfo{author}{\bibfnamefont{F.}~\bibnamefont{Behroozi}},
  \bibinfo{journal}{Langmuir} \textbf{\bibinfo{volume}{12}},
  \bibinfo{pages}{2289} (\bibinfo{year}{1996}).

\bibitem[{\citenamefont{Landau and Lifshitz}(1980)}]{la80}
\bibinfo{author}{\bibfnamefont{L.~D.} \bibnamefont{Landau}} \bibnamefont{and}
  \bibinfo{author}{\bibfnamefont{E.~M.} \bibnamefont{Lifshitz}},
  \emph{\bibinfo{title}{Statistical Physics}} (\bibinfo{publisher}{Pergamon},
  \bibinfo{address}{Oxford}, \bibinfo{year}{1980}), \bibinfo{edition}{3rd} ed.

\bibitem[{\citenamefont{Callen}(1985)}]{ca85}
\bibinfo{author}{\bibfnamefont{H.~B.} \bibnamefont{Callen}},
  \emph{\bibinfo{title}{Thermodynamics and an Introduction to
  Thermostatistics}} (\bibinfo{publisher}{John Wiley}, \bibinfo{address}{New
  York}, \bibinfo{year}{1985}).

\bibitem[{\citenamefont{Berny et~al.}(2000)\citenamefont{Berny, Rice, and
  Ross}}]{be00b}
\bibinfo{author}{\bibfnamefont{R.~S.} \bibnamefont{Berny}},
  \bibinfo{author}{\bibfnamefont{S.~A.} \bibnamefont{Rice}}, \bibnamefont{and}
  \bibinfo{author}{\bibfnamefont{J.}~\bibnamefont{Ross}},
  \emph{\bibinfo{title}{Physical Chemistry}} (\bibinfo{publisher}{Oxford
  University Press}, \bibinfo{address}{New York}, \bibinfo{year}{2000}).

\bibitem[{\citenamefont{Politano et~al.}(2012)\citenamefont{Politano, Marino,
  Campi, Far\'ias, Miranda, and Chiarello}}]{po12}
\bibinfo{author}{\bibfnamefont{A.}~\bibnamefont{Politano}},
  \bibinfo{author}{\bibfnamefont{A.~R.} \bibnamefont{Marino}},
  \bibinfo{author}{\bibfnamefont{D.}~\bibnamefont{Campi}},
  \bibinfo{author}{\bibfnamefont{D.}~\bibnamefont{Far\'ias}},
  \bibinfo{author}{\bibfnamefont{R.}~\bibnamefont{Miranda}}, \bibnamefont{and}
  \bibinfo{author}{\bibfnamefont{G.}~\bibnamefont{Chiarello}},
  \bibinfo{journal}{Carbon} \textbf{\bibinfo{volume}{50}}, \bibinfo{pages}{4903
  } (\bibinfo{year}{2012}).

\bibitem[{\citenamefont{Politano and Chiarello}(2015)}]{po15}
\bibinfo{author}{\bibfnamefont{A.}~\bibnamefont{Politano}} \bibnamefont{and}
  \bibinfo{author}{\bibfnamefont{G.}~\bibnamefont{Chiarello}},
  \bibinfo{journal}{Nano Res.} \textbf{\bibinfo{volume}{8}},
  \bibinfo{pages}{1847} (\bibinfo{year}{2015}).

\end{thebibliography}
\end{document}